\documentclass[showpacs,showkeys,preprint, preprintnumbers,amsmath,amssymb,nofootinbib, prd, aps, 11pt]{revtex4-1}

\usepackage{graphicx}
\usepackage{bbm}
\usepackage[T1]{fontenc}
\usepackage[latin1]{inputenc}
\usepackage{bezier, epsfig}
\usepackage{mathrsfs}
\newcommand{\Det}{\mbox{Det}}
\newcommand{\tr}{\mbox{tr}}
\newcommand{\Tr}{\mbox{Tr}}

\newcommand{\vek}[1]{\mathbf{#1}}

\newcommand{\ww}{\omega}
\newcommand{\wb}{\bar{\omega}}
\newcommand{\Ac}{\mathcal{A}}
\newcommand{\As}{\mathscr{W}}
\newcommand{\Asb}{\overline{\mathscr{W}}}
\newcommand{\chf}{\chi_{\rm fin}}

\renewcommand*{\d}[1][]{\mathop{\mathrm{d}^{#1}}\mkern-4mu}

\begin{document}


\title{A covariant variational approach to Yang-Mills Theory}

\author{M.~Quandt}\email{markus.quandt@uni-tuebingen.de}
\author{H.~Reinhardt}\email{hugo.reinhardt@uni-tuebingen.de}
\author{J.~Heffner}\email{jan.heffner@uni-tuebingen.de}
\affiliation{%
Institut f\"ur Theoretische Physik\\
Auf der Morgenstelle 14\\
D-72076 T\"ubingen\\
Germany
}%

\date{\today}


\begin{abstract}
We investigate the low-order Green's functions of $SU(N)$ Yang-Mills 
theory in Landau gauge, using a covariant variational principle based on the 
effective action formalism. Employing an approximation to the Faddeev-Popov 
determinant established previously in the Hamiltonian approach in Coulomb gauge
leads to a closed set of integral equations for the ghost and gluon propagator.
We carry out the renormalization and the infrared analysis of this system of equations. 
Finally, we solve the renormalized system numerically and compare with lattice results 
and other functional approaches. 
\end{abstract}


\maketitle


\section{Introduction}
\label{intro}
The low-order Green's functions of Yang-Mills theory have been the focus of many 
investigations, both in the continuum and on the lattice. Functional 
methods such as the functional renormalization group (FRG) flow equations 
\cite{[][{, and references herein}]Fischer:2008uz} and 
Dyson-Schwinger equations (DSE) \cite{[][{, and references herein}]Lerche:2002ep} initially 
concentrated their studies on the case of covariant gauges, since this is the natural choice 
in a covariant setup, and allows to use BRST symmetry and the ensuing Slavnov-Taylor identities 
to guide and improve the analysis. Moreover, the Kuga-Ojima criterion
\cite{Ojima:1978hy,Kugo:1979gm} argues that there should be a direct link between the 
deep infrared behavior of the gluon and ghost propagator, and the issue of colour 
confinement. Most investigations initially found an infrared vanishing, \emph{scaling} type of
solution for the gluon propagator, which is, however, at odds with high-precision lattice 
simulations \cite{Cucchieri:2007md, Cucchieri:2007rg, Bogolubsky:2009dc,Sternbeck:2013zja}. 
It was only later realized that infrared finite so-called \emph{decoupling} solutions could also 
be obtained in the functional approach under certain circumstances \cite{Fischer:2008uz}. 
In fact, an infrared finite gluon propagator had been found before in 
Refs.~\cite{Aguilar:2007nf,Binosi:2007pi,Aguilar:2008xm,Boucaud:2006if,Boucaud:2008ji,
Boucaud:2008ky,Dudal:2005na,Dudal:2007cw,Capri:2007ix,Dudal:2008sp,Oliveira:2007dy,Oliveira:2008uf}

By contrast, the so-called \emph{variational approach} to the Hamiltonian formulation of Yang-Mills theory
investigated in \cite{Schutte:1985sd, Szczepaniak:2001rg, 
Feuchter:2004mk,Reinhardt:2004mm, Epple:2006hv} has been formulated in \emph{Coulomb gauge}.
 This non-covariant condition is advantagous in the Hamiltonian approach since it allows for an 
explicit resolution of Gau\ss's law. In the Hamiltonian approach of \cite{Schutte:1985sd,Szczepaniak:2001rg,
Feuchter:2004mk,Reinhardt:2004mm,Epple:2006hv} the Schr\"odinger equation for the Yang--Mills vaccum wave 
functional is approximately solved by the variational principle, minimizing the energy with suitable 
trial \emph{Ans\"atze} for the vacuum wave functional. Compared to functional methods based on the 
Lagrangian formulation of quantum 
field theory, like the FRG \cite{Fischer:2008uz} and DSE \cite{Lerche:2002ep}, the variational principle 
of the Hamiltonian approach has the benefit that the relative size of the energy density controls the 
quality of the approximation made, i.e. of the trial wave functional used. By enlarging the space of 
trial states the description can be improved. The Hamiltonian approach in Coulomb gauge provides 
direct access to the so-called Coulomb potential between static charges and gives relatively simple 
explanations for confinement \cite{Reinhardt:2008ek,Reinhardt:2007wh,Pak:2009em,Reinhardt:2012qe,Reinhardt:2013iia} 
and other low-energy phenomena \cite{Pak:2011wu}. Furthermore it yields propagators which are in 
good agreement with lattice calculations \cite{Burgio:2008jr, Burgio:2009xp,Nakagawa:2009zf,
Burgio:2012bk} and the Gribov-Zwanziger confinement scenario \cite{Gribov:1977wm,Zwanziger:1991gz}. 
Unfortunately it cannot be generalized directly to covariant gauges and make contact with the alternative
studies mentioned above.

In the present paper, we will demonstrate that a conceptually similar 
variational principle can be established  in covariant gauges if one 
relies on the effective action instead of the energy. We will present a variational 
approach to quantum field theory which is based on the minimization of the effective action 
and apply it to Yang-Mills theory in Landau gauge. This approach will result in a system of 
integral equations for the low-order Green's functions, which we solve using the techniques 
borrowed from the Coulomb gauge calculations 
mentioned above. We will also argue that our method gives optimal results for the propagators
of the theory, while the extension to realistic vertices presumably requires to go beyond the 
Gaussian \emph{Ansatz} using Dyson-Schwinger equation techniques \cite{Campagnari:2010wc};
see also Ref.~\cite{Siringo:2013kua} for an alternative approach. 

The paper is organized as follows: 
In the next section, we present the general variational principle for the effective 
action and explain in section \ref{sec:3} how it can be applied to Yang-Mills theory in covariant 
gauges. After presenting our ansatz for the trial path integral measure,
we discuss an approximation to the Faddeev-Popov determinant which was introduced 
in Ref.~\cite{Reinhardt:2004mm} in the context of the Hamiltonian approach in Coulomb gauge 
and which facilitates the treatment of the ghost sector. In section \ref{sec:4} we apply the 
variation principle to the effective action of Yang-Mills theory in Landau gauge and derive a 
closed set of integral equations for the ghost and gluon propagator. The 
rather cavalier method to renormalize these equations is put on a solid ground in section \ref{sec:5}
by relating it to the traditional introduction of counter terms. After carrying out the 
infrared analysis of our renormalized system of equations, we present in section \ref{sec:6} 
their numerical solution and compare with recent high-precision lattice data. 
Finally, we give a short summary and end with an outlook on further developments.


\section{The variational principle}
\label{sec:2}
Below we recall the variational principle for the 
effective action in quantum field theory. This material mainly serves to fix some of 
our notation and introduces various concepts used lateron.
We start with a theory for a quantum field $\phi$ defined by an action $S(\phi)$ 
in $d=n+1$ dimensional Euclidean space time. Expectation values of operators are 
computed with the normalized measure
\begin{eqnarray}
d\mu_0(\phi) &=& Z^{-1}\,\exp\left[ - \hbar^{-1}\,S(\phi)\right]\,d\phi 
\label{1.1}
\\[2mm]
Z &\equiv& \int d\phi\,\exp\left[ - \hbar^{-1}\,S(\phi)\right]\,.
\nonumber
\end{eqnarray}
Here $d\phi$ is the flat (translationally invariant) measure in field space, 
and an implicit regularization is understood. The obvious analogy 
with statistical mechanics can now be exploited to define a variation method: 
The Gibbs-like measure eq.~(\ref{1.1}) is the unique 
probability measure in field space which minimizes the free energy (or rather, 
free action),
\begin{equation}
F(\mu) \equiv \langle S\rangle_\mu - \hbar \As(\mu) 
\stackrel{!}{=} \mbox{min}\,.
\label{1.2}
\end{equation}
Here $\langle \ldots \rangle_\mu \equiv \int d\mu(\phi)\ldots$ is 
the expectation value in the trial measure $\mu$, and the normalization is such that
$\langle 1 \rangle_\mu =  \int d\mu(\phi) = 1$.
If the measure is written in Radon-Nikodym form $d\mu(\phi) = d\phi\,\rho(\phi)$ 
with a suitable density $\rho$, the quantity
\begin{equation}
\As(\mu) \equiv - \langle \ln \rho \rangle_\mu = 
- \int d\phi\,\rho(\phi)\,\ln \rho(\phi)
\label{1.3}
\end{equation}
is the \emph{entropy} which measures the accessible field space for 
quantum fluctuations. The value of the minimal free action 
from eq.~(\ref{1.2}) attained for the Gibbs measure (\ref{1.1}) is 
$F(\mu_0) = - \hbar\,\ln Z$.
Notice that Planck's constant really plays a role similar to the
temperature in statistical mechanics, i.e.~it controls the balance between the
classical action $\langle S\rangle$ and the fluctuation entropy $\As(\mu)$.

Traditionally, one would now take the minimising measure eq.~(\ref{1.1}), or some 
approximate minimum in a restricted measure space, and compute the Schwinger functions 
\begin{equation}
G_n(x_1,\ldots,x_n) \equiv \langle \phi(x_1)\cdots\phi(x_n)\rangle
= \int d\mu_0(\phi)\,\phi(x_1)\cdots \phi(x_n)\,,
\label{1.4}
\end{equation}
or rather the generating functional $W(j)$ of their connected part. 
From this, the effective action $\Gamma$ follows 
by Legendre transformation.

Alternatively, the effective action can also be characterized directly by a 
variation principle: To this end, we go back to eq.~(\ref{1.2}) and perform 
the minimization in two steps: 
First, we define a constrained free action by restricting some operator
$\Omega(\phi)$ to a classical value $\omega$, and second, we minimise this
action under variation of all trial probability measures, which yields the 
\emph{effective action} for the operator $\Omega$,
\begin{equation}
\Gamma(\omega) \equiv  \min_\mu F(\mu,\omega) \equiv \min_\mu 
\Big\{ \langle S\rangle_\mu - \hbar \As(\mu) \,\big|\,
\langle \Omega \rangle_\mu = \omega  \Big\}\,.
\label{1.5}
\end{equation}
It depends, of course, on the choice of the operator $\Omega$, and is a functional 
of the prescribed value $\omega$ both explicitly (via the constraint) and implicitly 
(via the $\omega$-dependence of the solving measure). In most cases, the constraint 
is chosen as the vev of the quantum field itself, so that
\begin{equation}
\Gamma(\varphi) \equiv \min_\mu\,\Big\{  \langle S\rangle_\mu - \hbar \As(\mu) \,\big|\,
\langle \phi \rangle_\mu = \varphi \, \Big\}\,.
\label{1.6} 
\end{equation}
The variation principle eq.~(\ref{1.2}) is now equivalent to $\Gamma(\varphi) 
 \stackrel{!}{=} \mbox{min}$, which is a problem in classical field theory that can 
be considered as solved. We now have two distinct definitions of the effective action:
\begin{itemize}
 \item[\textbf{(i)}] \emph{Functional:} Define the Gibbs measure as solution of 
eq.~(\ref{1.2}), compute the Green functions, or their generating functional,  
from eq.~(\ref{1.4}), and finally construct the effective action as generating functional of 
1PI correlators by means of a Legendre transformation.
\item[\textbf{(ii)}] \emph{Linear response:} Compute the effective action directly from  
the variation principle eq.~(\ref{1.6}) and obtain the 1PI proper $n$-point functions 
as derivatives at $\varphi=0$. 
\end{itemize}
It is not hard to see that the two descriptions agree, cf.~appendix \ref{app:A}, but 
this identity only holds for the \emph{exact} solution of the variation problem, 
when no restrictions are placed on the trial measures $\mu$. This is rarely ever the case. 
In practice, a viable variation scheme will have to restrict the space of trial 
measures $\{ \mu \}$ to those candidates for which the expectation values in 
eq.~(\ref{1.2}) or eq.~(\ref{1.6}) can be computed explicitly. In such a restricted 
variation, the effective action determined from \textbf{(i)} and \textbf{(ii)} will 
differ. In appendix \ref{app:B}, we sketch the two approaches 
for a simple $\phi^4$ theory with Gaussian trial measures, and also compare with the 
variational solution of the Schr\"odinger equation in the Hamilton formalism. 

As with all variational methods, it is not \emph{a priori} clear which of the two 
formulations above will give the better approximation to the true system, although 
general arguments \cite{Parisi:1988nd} indicate that the linear response approach 
\textbf{(ii)} is of higher order in the difference between true and approximate 
minimal measure $\mu_0$. (However, higher order does not always mean higher accuracy.)
In any case, the representation eq.~(\ref{1.6}) is conceptually simpler and automatically
ensures that $\Gamma(\varphi)$ is a convex \emph{upper bound} to the true 
effective action whenever restrictions are placed on the measure $d\mu(\phi)$. 


\section{Application of the variational principle to Yang-Mills theory in covariant gauges}
\label{sec:3}

Since our variation principles are covariant, it is natural to study $SU(N)$
Yang-Mills theory in covariant gauges, in particular Landau gauge. 
Ignoring Gribov copies, the exact measure for this problem is
\begin{eqnarray}
d\mu_0(A) &=& Z^{-1}\,\mathscr{J}(A)\,\exp\big[ - \hbar^{-1}\,S_{\rm gf}(A) \big] d A\\[2mm]
\label{3.0}
Z &=& \int dA \, \mathscr{J}(A)  \exp\left[ - \hbar^{-1}\,S_{\rm gf}(A)\right]\nonumber \\[2mm]
S_{\rm gf} &=&  \frac{1}{2} \,\|F_A\|^2 +
\frac{1}{2\xi} \,\|d^\dagger A\|^2 \,,
\label{3.1}
\end{eqnarray}
where $A$ and $F_A$ are the differential forms for the gauge connection and 
its field strength, respectively,
\begin{eqnarray}
A &=& A_\mu \,dx^\mu = A_\mu^a T^a\,dx^\mu \nonumber \\[1mm]
F_A &=& dA + g\, A \wedge A = \frac{1}{2} \,F_{\mu\nu}\,dx^\mu \wedge dx^\nu =
\frac{1}{2} \,F^a_{\mu\nu}T^a\,dx^\mu \wedge dx^\nu\,,
\end{eqnarray}
and $g$ is the bare coupling strength.
As usual, Feynman gauge ($\xi=1$) simplifies the Lorentz structure of the propagator, while 
Landau gauge ($\xi=0$) yields transversal gluons that can be compared directly to lattice 
investigations.  
The prefactors in the action and the inner product of Lie-Algebra valued forms on 
4D euclidean space $M$, 
\begin{align*}
(\eta,\omega) \equiv (-2)\,\tr\,\int_M \eta \wedge \ast\omega  = 
\int_M \eta^a \wedge \ast\omega^a
\end{align*}
are consistent with \emph{antihermitean}
generators of the Lie--algebra, normalised according to 
$\tr \,T^a T^b = - \frac{1}{2} \delta ^{ab}$. Moreover,
the measure factor $\mathscr{J}(A)$ in eq.~(\ref{3.1}) is the (normalized) Faddeev--Popov
determinant
\begin{equation}
\mathscr{J}(A) \equiv \Det \left[ - \partial_\mu\hat{D}_\mu^{ab} \right]
/ \Det \left[ - \Box\,\delta^{ab}\right]
= \Det\,\left[ - \Box\,\delta^{ab} - g \,\partial_\mu f^{abc} A_\mu^c 
\right] / \Det \left[ - \Box\,\delta^{ab}\right]
\label{3.2}
\end{equation}
with $\mathscr{J}(0) = 1$. It is helpful to interpret $\mathscr{J}(A)$ as the weight of 
the gauge orbit through $A$, i.e.~the canonical volume form on this orbit in field space. 

\subsection{Modified variational principle}
The appearance of the Faddeev-Popov determinant in the measure eq.~(\ref{3.0}) requires some 
modification of the basic variation principle eq.~(\ref{1.2}), because the latter only 
holds for measures of the Gibbs form eq.~(\ref{1.1}). One obvious solution is to transfer
the Faddeev-Popov determinant into the action, 
\[
S_{\rm gf} \to \bar{S} = S_{\rm gf} - \hbar \ln \mathscr{J}\,.
\]
Then the variation principle based on $\bar{S}$ takes the standard form
$F(\mu) \stackrel{!}{=} \mathrm{min}$, where 
\begin{equation}
F(\mu) = \langle \bar{S}\rangle_\mu - \hbar \As(\mu) = 
\langle S_{\rm gf} \rangle_\mu - \hbar \,\langle \ln\mathscr{J}(A)\rangle_\mu
+ \hbar \,\langle \ln \rho(A) \rangle_\mu \,.
\label{3.3}
\end{equation}
Here, $\rho$ is the deviation of the trial measure from the 
flat measure (excluding the Faddeev-Popov determinant), $d\mu = dA\, \rho(A)$. 
The rhs. of eq.~(\ref{3.3}) suggests to rewrite the 
variation principle by redefining the entropy, 
\begin{eqnarray}
F(\mu) &=& \langle S_{\rm gf}\rangle_\mu - \hbar\,\Asb(\mu)
\nonumber \\[2mm]
\Asb(\mu) &\equiv& \As(\mu) + \left\langle \ln( \mathscr{J}
) \right\rangle_\mu =- \left\langle \ln(\rho / \mathscr{J}
) \right\rangle_\mu = - \langle \ln \bar{\rho}\rangle_\mu\,.
\label{3.4}
\end{eqnarray}
Here, $\bar{\rho}$ is now the deviation from the standard measure 
\emph{including} the Faddeev-Popov determinant, which is the natural metric on the 
space of gauge orbits, 
\begin{equation}
  d\mu = dA\, \rho(A) = dA\,\mathscr{J}(A)\, \bar{\rho}(A)\,.
\end{equation}
The redefined entropy eq.~(\ref{3.4}) coincides with the usual notion of the 
\emph{relative entropy} of the trial volume form $dA\,\rho(A)$ 
compared to the standard weight $dA\mathscr{J}(A)$ on the space of gauge orbits. 
As a consequence, the general variational approach sketched above remains valid for 
YM theory in covariant gauges, if we only replace the entropy $\As$ by the 
\emph{relative} entropy $\Asb$.

\subsection{Gaussian trial measure}
In the next step, we have to choose a class of suitable trial measures which is simple
enough to allow for the necessary integrals to be performed, but still
captures the essential physics of the system. For this purpose, we note
\begin{itemize}
\item[\textbf{(i)}] gluons are only weakly interacting in the ultra-violet due to
asymptotic freedom
\item[\textbf{(ii)}] gluon configurations near the Gribov horizon ($\mathscr{J}=0$) are 
assumed to play a dominant role in the infrared, and the self-interactions
of gluons in such configurations may become sub-dominant 
\end{itemize}
The overall picture is \textbf{(i)} an (almost) non-interacting constituent gluon with 
\textbf{(ii)} an enhanced weight near the Gribov horizon. It is precisely this picture which is supported by the variational calculation in the Hamiltonian approach to Yang-Mills theory in Coulomb gauge developed in Refs.~\cite {Reinhardt:2004mm,Feuchter:2004mk,Epple:2006hv}. The first condition implies that the 
trial action should be (close to) Gaussian, while the enhancement at the horizon is controlled 
by the volume form on the gauge orbit: Since the natural volume form is $\mathscr{J}(A)$, 
replacing it in the trial measure by $\mathscr{J}^{1- 2\alpha}$ with $\alpha \ge 0$ will enhance 
the weight of near-horizon configurations by a relative factor $\mathscr{J}(A)^{-2\alpha} \gg 1$. 

\medskip
We will thus attempt a variational approach based on trial measures of the form 
($\hbar= 1$)
\begin{equation}
d\mu(A) = \mathscr{N}_\alpha\cdot \,dA \, \mathscr{J}(A)^{1-2\alpha} 
\,\exp\left[ - \frac{1}{2} \,\int d^4(x,y)\,A_\mu^a(x)\,
\omega_{\mu\nu}^{ab}(x,y)\,A_\nu^b(y) \right],
\label{3.5}
\end{equation}
where $\mathscr{N}_\alpha$ is the overall normalisation.\footnote{The normalisation factor 
$\mathscr{N}_\alpha$ will, in general, depend on the variational kernel $\omega$. In 
the case $\alpha = \frac{1}{2}$, for instance, we have 
$\mathscr{N}_{\frac{1}{2}} = \mathrm{det}\big[\omega / (2 \pi)\big]^{\frac{1}{2}}$.} 
Note that for $\alpha = 0$ the Gaussian represents the relative weight $\bar{\rho}(A)$ while for 
$\alpha = \frac{1}{2}$ it gives the full weight $\rho(A)$, cf.~eq.~(\ref{3.4}). 
We will treat $\alpha$ as a variational parameter, although we shall find below that the exact 
value of $\alpha$ is immaterial, at least up to two loop order in a formal loop counting scheme
introduced in the next subsection. 

\medskip
The measure (\ref{3.5}) is unconstrained and thus appropriate for the functional approach 
discussed in section \ref{sec:2}. (We present the necessary modifications to comply with 
the constraint $\langle A \rangle = \mathcal{A}$ in eq.~(\ref{3.13}) below.)  
In the absence of an external classical field $\mathcal{A}$  the variational method 
maintains global color and Lorentz symmetry, and the kernel $\omega$ can be chosen diagonal 
and transversal up to the covariant gauge fixing term from eq.~(\ref{3.1}),
\begin{eqnarray}
\omega^{ab}_{\mu\nu}(x,y) &=& \int \frac{\d^d k}{(2 \pi)^4}\,
e^{i k(x-y)}\,\ww^{ab}_{\mu\nu}(k)
\nonumber \\[2mm]
\ww^{ab}_{\mu\nu}(k) &=& \delta^{ab}\,\left[\delta_{\mu\nu} -
\frac{k_\mu k_\nu}{k^2}\, (1 - \xi^{-1}) \right]\,\ww(k)
\equiv \delta^{ab}\,t_{\mu\nu}(k)\,\ww(k)\,.
\label{3.6}
\end{eqnarray}
The normalisation in eq.~(\ref{3.5}) is such that the bare gluon propagator 
\begin{align}
D^{ab}_{\mu\nu}(x,y) \equiv \langle A_\mu^a(x)\,A_\nu^b(y)\rangle 
= \int \frac{\d^d k}{(2\pi)^d}\,e^{i k (x-y)}\,D^{ab}_{\mu\nu}(k)
\label{3.6.1}
\end{align}reduces for $g \to 0$ (and hence $\mathscr{J} \to 1$) to
\begin{align}
D^{ab}_{\mu\nu}(k) \stackrel{\mathscr{J} \to 1}{\longrightarrow} 
\left[\ww(k)^{-1}\right]^{ab}_{\mu\nu} = \delta^{ab}\,t_{\mu\nu}^{-1}(k)\,\frac{1}{\ww(k)}
\stackrel{g \to 0}{=} \delta^{ab}\,t_{\mu\nu}^{-1}(k)\,\frac{1}{k^2}\,,
\label{3.6.2}
\end{align}
i.e.~$\ww(k) \to k^2$. This is also the UV limit of $\omega(k)$ due to asymptotic freedom.
Here, the Lorentz structure is given by the inverse 
\begin{equation}
t^{-1}_{\mu\nu}(k) = \delta_{\mu\nu} - \frac{k_\mu k_\nu}{k^2}\,(1-\xi) \,;
\label{3.7}
\end{equation}
for Landau gauge ($\xi = 0$) this becomes the transversal projector.  

\medskip
A possible caveat against the trial measure eq.~(\ref{3.5}) is that it does not 
respect the BRST symmetry of the full theory, nor any of the identities that follow from it. 
This is, in a sense, unavoidable in a variational ansatz, because the simplest 
non-topological action with full BRST symmetry is already the full YM theory, so that
\emph{any} truncation will necessarily break the Slavnov-Taylor identities to a 
certain extent. For Landau gauge, however, recent lattice calculations favour a 
soft BRST breaking massive gluon propagator in the deep infrared, and such 
\emph{decoupling} solutions were also found under certain assumptions within 
functional approaches \cite{Aguilar:2007nf,Binosi:2007pi,Aguilar:2008xm,Boucaud:2006if,Boucaud:2008ji,Boucaud:2008ky,Dudal:2005na,Dudal:2007cw,Capri:2007ix,Dudal:2008sp}. Since a dynamical mass generation is one of the main virtues 
of variational methods, the ansatz eq.~(\ref{3.5}) seems therefore justified for 
covariant gauges.

\subsection{Curvature approximation}
The Faddeev-Popov determinant eq.~(\ref{3.2}) and its expectation value in the trial 
measure (\ref{3.5}) cannot be computed in closed form. In the following, we will adopt 
an approximation that has been shown to be correct up to two-loop order in 
the energy functional within the variational Hamiltonian approach \cite{Reinhardt:2004mm} 
in \emph{Coulomb gauge}:
Since $(\ln \mathscr{J})$ and
$\delta \ln \mathscr{J} / \delta A$ both vanish at $A=0$, we can write
\begin{equation}
\ln \mathscr{J}[A] = -\frac{1}{2}\int d(x,y) \,K^{ab}_{\mu\nu}(x,y)\,A_\mu^a(x)\,A_\nu^b(y)
\label{3.8}
\end{equation}
with a symmetric kernel $K[A]$ that may depend arbitrarily on the gauge connection $A$.
As a consequence, 
\begin{align}
K(1,2) = - \frac{\delta^2 \ln \mathscr{J}}{\delta A(1)\,\delta A(2)} - 
\left \{ \frac{\delta K(1,3)}{\delta A(2)} + \frac{\delta K(2,3)}{\delta A(1)}\right \}\,A(3)
- \frac{1}{2}\,\frac{\delta K(3,4)}{\delta A(1)\,\delta A(2)}\,A(3)\,A(4)\,, 
\label{3.8a}
\end{align}
where each digit stands for the combination of colour, Lorentz and spacetime indices, and 
repeated indices are summed or integrated over. If we now \emph{formally} introduce a loop 
counting parameter in the exponent of the trial measure (\ref{3.5}), we find from eqs.~(\ref{3.8})
and (\ref{3.8a})
\begin{align*}
\langle  \, \ln \mathscr{J}[A]\,\rangle &= \left \langle \frac{1}{2}\, \frac{\delta^2 \,
\ln \mathscr{J}}{\delta A(1)\,\delta A(2)}\,A(1) A(2)\right\rangle
+ \left \langle \frac{1}{2}\, \left \{ \frac{\delta K(1,3)}{\delta A(2)} + \frac{\delta K(2,3)}{\delta A(1)}\right \}\,
A(1) A(2) A(3) \right\rangle  + \nonumber \\[2mm]
&\quad + \left\langle \frac{1}{4}\,\frac{\delta K(3,4)}{\delta A(1)\,\delta A(2)}\,A(1) A(2) A(3) A(4) \right \rangle
\nonumber \\[2mm]
&=
\left \langle \frac{1}{2}\, \frac{\delta^2 \ln \mathscr{J}}{\delta A(1)\,\delta A(2)}\right \rangle\,
\big\langle A(1) A(2) \big\rangle + \cdots\,, 
\end{align*}
where the expectation value is taken with the trial measure, eq.~(\ref{3.5}), and the dots indicate 
higher orders in the loop counting scheme. To leading loop order, we can therefore 
replace the kernel $K[A]$ by the average curvature on the gauge 
orbit\footnote{Besides being defined in $d=4$ Euclidean dimensions the curvature $\chi$ introduced 
in eq.~(\ref{3.9}) differs from the one defined in Ref.~\cite{Reinhardt:2004mm} by a factor of $2$; the 
same is true for the kernel $\omega$.},  
\begin{align}
K^{ab}_{\mu\nu}(x,y) \,\,\,\to\,\,\,  \chi^{ab}_{\mu\nu}(x,y) \equiv 
-\,\left\langle \frac{\delta^2 \ln \mathscr{J}}{\delta A_\mu^a(x)\,\delta A_\nu^b(y)}
\right \rangle\,.
\label{3.9}
\end{align}
In total, this approximation then yields the formula \cite{Reinhardt:2004mm}
\begin{align}
\ln \mathscr{J}[A] \approx - \frac{1}{2}\int d(x,y)\,\chi^{ab}_{\mu\nu}(x,y)\cdot A_\mu^a(x)\,A_\nu^b(y)
\label{3.10} 
\end{align}
which differs, on average, from the exact expression (\ref{3.8}) only by a higher loop 
effect.\footnote{In Ref.~\cite{Reinhardt:2004mm}, it was shown by explicit calculation that
eq.~(\ref{3.10}) is, in fact, exact up to including two loops in the effective action.}
However, the omitted higher loop terms need not be negative definite and we 
cannot guarantee that our approximate effective action is always a strict upper bound to the 
true effective action. We will refer to eq.~(\ref{3.10}) in the following as the
\emph{curvature approximation}. 

\medskip
The salient point of this approximation is now that the curvature eq.~(\ref{3.9}) is easier to 
compute than the expectation value of the full Faddeev-Popov determinant. As shown in section \ref{app:chi},
the curvature can be related to the ghost propagator (although we do not explicitly introduce ghosts),
which in turn can be evaluated from a Dyson equation that involves the kernel $\omega$ and the 
full ghost-gluon vertex. In the rainbow-ladder approximation, this full vertex is replaced by the 
bare one and $\chi$ becomes a well-defined function of the ghost propagator and the variation 
parameters $\omega$ and $\alpha$; for further details cf.~section \ref{app:chi}.
This approximation is further supported by the fact that the ghost-gluon vertex in 
Landau gauge is not renormalized \cite{Taylor:1971ff} and shows little dressing in 
lattice simulations \cite{Cucchieri:2004sq}, i.e.~the radiative corrections 
to the vertex tend to cancel.

\bigskip
We can now use the curvature approximation eq.~(\ref{3.10}) directly in the trial measure (\ref{3.5}).
From global colour and Lorentz symmetry of our variational ansatz (in the functional formulation 
without an external classical field $\mathcal{A}$), the expectation value in the definition eq.~(\ref{3.9}) 
entails that the curvature has the same simple Lorentz structure as the variation kernel
\begin{align}
\chi^{ab}_{\mu\nu}(x,y) = \delta^{ab}\,t_{\mu\nu}(k)\,\chi(k)\,, 
\label{3.11}
\end{align}where $\chi(k)$  is known as the \emph{scalar curvature}. 
The trial measure eq.~(\ref{3.5}) thus depends on the curvature and the parameter $\alpha$ 
only in the combination
\begin{equation}
\wb(k) \equiv \ww(k) + (1- 2 \alpha)\,\chi(k)\,.
\label{3.12}
\end{equation}
This has the same effect as putting $\alpha = \frac{1}{2}$ and replacing $\ww \to \wb$
in eq.~(\ref{3.5}), i.e.~the Faddeev-Popov determinant drops out from our variational 
ansatz within the curvature approximation which then becomes a simple Gaussian with kernel
$\wb(k)$, and hence results in the gluon propagator (cf.~eq.~(\ref{3.6.1})),
\begin{align}
D^{ab}_{\mu\nu}(k) = \delta^{ab}\,t_{\mu\nu}^{-1}(k)\,\frac{1}{\wb(k)}\,.
\label{G10} 
\end{align}
This observation greatly simplifies the computation of expectation values, 
since we then have \emph{Wick's theorem} at our disposal. It must be stressed, however, 
that the variation is still with respect to $\ww(k)$, \emph{not} $\wb(k)$, because the 
curvature $\chi(k)$ is, in principle, a dependent quantity. 
 
\bigskip
So far, we have mainly discussed the unconstrained measure for the functional approach, 
when no external classical field $\mathcal{A}$ is prescribed. As discussed in appendix 
\ref{app:B}, this is entirely sufficient to investigate the \emph{propagators}, 
since these agree in the functional and linear response approach. 
Although we will not study the non-trivial vertex corrections arising in the 
linear response formulation in any detail, we still want to show at least how the variation 
problem can be set up in this case: First, we have to adjust the trial measures introduced 
above to comply with the constraint $\langle A \rangle = \mathcal{A}$ imposed by the classical 
field. This can always be achieved by \emph{shifting} the gauge field $A_\mu \to A_\mu - \mathcal{A}_\mu$ 
in the (full) density $\rho(A)$ of the trial measure.\footnote{This statement is not restricted to 
Gaussian measures: since the full density $\rho(A)$ multiplies, by definition, the flat measure $dA$, 
the proposed shift in $\rho(A)$ leads to  
\begin{align}
\langle A \rangle = \int dA\,\rho(A)\cdot A \longrightarrow \int dA\,\rho(A-\mathcal{A})\cdot 
A \stackrel{(*)}{=} \int dA\,\rho(A)\cdot A + \mathcal{A}\,\int dA\,\rho(A) = 
\langle A \rangle + \mathcal{A}\,, 
\end{align}
where we used the translation invariance of the flat measure in $(*)$. In the absence of a 
classical field, Lorentz invariance entails $\langle A \rangle = 0$, and the constraint 
$\langle A \rangle = \mathcal{A}$ follows.
} 
In the present case, the density $\rho(A)$ is Gaussian after applying the curvature 
approximation to eq.~(\ref{3.5}), so that the final form of our trial probability 
measure for the linear response approach is
\begin{align}
\rho(A) = \mathrm{Det}\,\left(\frac{\wb}{2 \pi}\right)^{\frac{1}{2}}
\,\exp\left[ - \frac{1}{2} \,\int d^4 x d^4 y\,
(A_\mu^a(x) - \Ac_\mu^a(x))\,\wb_{\mu\nu}^{ab}(x,y)\,
(A_\nu^b(y)-\Ac_\nu^b(y)) \right]\,.
\label{3.13}
\end{align}
It should be emphasized again that the optimal kernel $\wb_\mathcal{A}$ determined from 
eq.~(\ref{3.13}) will depend implicitly on the classical 
field $\mathcal{A}$, which is externally prescribed and thus arbitrary. As a consequence, 
we can no longer assume the simple colour and Lorentz structure (\ref{3.6}) valid in the 
functional approach; instead we would have to deal with the full matrix gap equation in 
position space, cf.~appendix \ref{app:X}.

\section{The effective action of Yang-Mills theory}
\label{sec:4}

We are now in a position to determine the effective action of Yang-Mills theory from 
the variational principle using the trial probability measure (\ref{3.13}).

\subsection{The free action}
\label{sec:F}
To evaluate the free action of the trial measure (\ref{3.13}), we first
expand the classical Yang-Mills Lagrangian including the gauge fixing term,
\[
\mathscr{L}_{\rm gf} = \frac{1}{2} 
A_\mu^a \big[ - \Box \delta_{\mu\nu} +
(1-\xi^{-1})\,\partial_\mu\partial_\nu \big]
A_\nu^a + g\, f^{abc} (\partial_\mu A_\nu^a) A_\mu^b A_\nu^c +
\frac{g^2}{4} f^{abc} f^{ade} A_\mu^b A_\nu^c A_\mu^d A_\nu^e\,.
\]
The relevant correlators in the measure eq.~(\ref{3.13}) can easily be 
worked out using Wick's theorem,
\begin{align}
\langle\,A_\mu^a(x)\,\rangle_\mu = &\,\,\Ac_\mu^a(x)  \nonumber \\[1mm]
\langle\,A_\mu^a(x)\,A_\nu^b(y)\,\rangle_\mu = &\,\,\Ac_\mu^a(x)\,\Ac_\nu^b(y)
+  [\wb^{-1}]_{\mu\nu}^{ab}(x,y)  \nonumber\\[1mm]
\langle\,A_\mu^a(x)\,A_\nu^b(y)\,A_\alpha^c(z)\,\rangle_\mu = 
 &\,\left([\wb^{-1}]^{ab}_{\mu\nu}(x,y)\,\Ac^c_\alpha(z) + 2 \mbox{ perm.}
\right) + \Ac^a_\mu(x)\,\Ac^b_\nu(y)\,\Ac^c_\alpha(z) \nonumber\\[1mm]
\langle\,A_\mu^a(x)\,A_\nu^b(y)\,A_\alpha^c(z)\,A_\beta^d(u)\,\rangle_\mu =
 &\,\left( [\wb^{-1}]^{ab}_{\mu\nu}(x,y)\,[\wb^{-1}]^{cd}_{\alpha\beta}
(z,u) + 2 \mbox{ perm.} \right) +   \nonumber\\[1mm]
& {}+ \left( [\wb^{-1}]^{ab}_{\mu\nu}(x,y)\,\Ac_\alpha^c(z)\,
\Ac_\beta^d(u) + 5 \mbox{ perm.} \right)+  \nonumber\\[1mm]
& {} +  \Ac_\mu^a(x) \Ac_\nu^b(y)  \Ac_\alpha^c(z)  \Ac_\beta^d(u)\,.
\label{3.13.1}
\end{align}
Next, we have to insert this into $\langle\,S_{\rm gf}\,\rangle$
and combine it with the relative entropy (cf.~eq.~(\ref{3.17}) below) to obtain the 
free action $F(\omega,\mathcal{A})$. The resulting expressions are, however, rather 
complicated because the kernel $\wb$ in the linear response approach
does not have the colour and Lorentz symmetry eq.~(\ref{3.6}). On the other hand, 
this complication is unnecessary: as argued in appendix \ref{app:B}, it is sufficient
to use the functional approach with $\mathcal{A} = 0$ and the symmetric kernel 
eq.~(\ref{3.6}), as long as we are only interested in the propagators of the theory.
For completeness, we present the full expression for the free action 
$F(\omega,\mathcal{A})$ as well as the ensuing gap equation in appendix \ref{app:X}, 
but we do not investigate the resulting vertex corrections in more detail. 

\medskip
For the remainder of this paper, we therefore concentrate on the propagators, i.e.~we set 
$\mathcal{A} = 0$ and use translational invariance to transform to momentum space
as in eq.~(\ref{3.6}). The global colour symmetry $\wb^{ab} \sim \delta^{ab}$ 
combined with the anti-symmetry of the structure constants $f^{abc}$ then allows
to perform all colour traces by means of the the SU$(N)$ relations
\[
f^{abc}\,f^{abd} = N\,\delta^{cd} \quad \Longrightarrow \quad
f^{abc}\,f^{abc} = N(N^2-1) \,.
\]
The result for the average classical action is
\begin{align}
\langle\,S_{\rm gf}\,\rangle &= S_{\rm gf}(\Ac) + \frac{1}{2}\,V_d \, (N^2-1)\,b_0\int 
\frac{d^d k}{(2 \pi)^d}\,\frac{k^2}{\wb(k)}  +
\nonumber \\[2mm]
& \quad { } + \frac{g^2}{4}\,V_d\,N(N^2-1) \,\left \{ b_1 \,\Omega^2 - 
b_2\,\int \frac{d^d (p,k)}{(2 \pi)^{2d}}\,\frac{(p\cdot k)^2}{p^2\,k^2}\,
\frac{1}{\wb(p)\,\wb(k)} \right \}  
\,.
\label{3.14}
\end{align}
Here, $V_d$ is the spacetime volume and the numerical factors are
\begin{align}
b_0 &= d_\xi \nonumber \\[2mm]
b_1 &=  d^2 - 3d + 3 + 2 (d-2)\,\xi + \xi^2 \nonumber \\[2mm]
b_2 &= (1 - \xi)^2 
\,,
\label{3.16}
\end{align}
where the symbol $d_\xi$ in the first line means $d$ for all $\xi \neq 0$, and $(d-1)$ for $\xi=0$.
The kernel $\wb$ appears in eq.~(\ref{3.14}) both explicitly and also within 
the momentum-independent expressions
\begin{eqnarray}
\Omega_{\mu\nu} &\equiv& 
\int \frac{d^d k}{(2\pi)^d}\,\frac{k_\mu k_\nu}{k^2}\,\frac{1}{\wb(k)} 
\nonumber \\[2mm]
 \Omega &\equiv& \Omega_{\mu\mu} = \int \frac{d^d k}{(2 \pi)^d}\,\frac{1}{\wb(k)}\,.
\label{3.15}
\end{eqnarray}

Next we need the entropy of the Gaussian measure, which can be calculated in 
the same way as for the $\phi^4$ case (cf.~appendix \ref{app:B}). The additional 
colour indices on the kernel $\wb$ in the measure simply yield an overall factor of 
$N^2 -1$, so that
\begin{align}
\As &=  \frac{1}{2} \,\mathrm{Tr}\,\left\{\, \mathbbm{1} - \ln \left(\frac{\omega}{2 \pi\hbar}
\right)\,\right\} \nonumber \\[2mm]
&= - \frac{1}{2}\,(N^2-1)\, V_d\,\int\frac{d^d k}{(2\pi)^d}\,
\mathrm{tr}\,\ln \left[\left(\delta_{\mu\nu} - \frac{k_\mu k_\nu}{k^2} (1-\xi^{-1})
\right)\,\wb(k)\right] + \text{const}\,,
\label{3.17} 
\end{align}
where the trace $\mathrm{tr}$ is with respect to the Lorentz indices. 
For $\xi \neq 0$ we can now use the following identity for $(d\times d)$ matrices,
\begin{align}
\tr\, \ln (t_{\mu\nu}\cdot \wb) = \ln\det(\wb\cdot t_{\mu\nu}) = 
\ln(\wb^d\cdot \det t_{\mu\nu}) = d\,\ln \wb + \ln \det t_{\mu\nu} \,.
\label{3.18}
\end{align}
The last term is $\omega$-independent and may thus be dropped, so that\footnote{The 
formula also holds for Landau gauge $\xi=0$ if the factor of $d$ is replaced 
by $(d-1)$.} 
\begin{equation}
\As  = - (N^2-1)\,\frac{1}{2}\,d_\xi \, V_d \int \frac{d^d k}{(2\pi)^d}\,
\ln \wb(k) + \mathrm{const}\,.
\label{3.19}
\end{equation}
This is the full entropy of the measure eq.~(\ref{3.13}) with $\mathcal{A} = 0$ and the 
symmetric kernel $\wb$. As pointed  out earlier, the free action in the YM case must, 
however, be based on the \emph{relative} entropy $\Asb$ eq.~(\ref{3.4}), which differs 
from eq.~(\ref{3.19}) by the expectation value of the Faddeev-Popov determinant
\begin{equation}
\Asb = - (N^2-1)\,\frac{1}{2} \, d_\xi \, V_d 
\int \frac{d^d k}{(2\pi)^d}\,\ln \wb(k) +\langle \ln \mathscr{J} \rangle_{\wb}\,.
\label{3.20}
\end{equation}
We will later treat the last term in curvature approximation, but for now we
keep it general. Eventually, we find the free action as the difference between the 
average action eq.~(\ref{3.14}) and the relative entropy eq.~(\ref{3.20}), 
\begin{align}
F(\wb) =  \langle S_{\rm gf}\rangle_\omega  - \Asb(\wb)\,.
\label{3.21}
\end{align}

\subsection{The gap equation}
\label{sec:gap}
Our task is to minimize the free action eq.~(\ref{3.21}) with respect to 
the kernel $\ww(k)$. From eq.~(\ref{3.12}) and the fact that the curvature $\chi(k)$ is, in 
principle, a $\ww$-dependent quantity, we have
\begin{align}
\frac{\delta}{\delta \ww(k) } = \frac{\delta}{\delta \wb(k)} + 
(1- 2 \alpha)\,\int dp\,\frac{\delta \chi(p)}{\delta \ww(k)}\cdot
\frac{\delta}{\delta \wb(p)}\,. 
\label{3.22}
\end{align}
The second term describes the implicit change of the curvature with the 
variation kernel. While an integral type of equation for this quantity can, 
in principle, be written down, it represents a higher order effect and 
will thus be neglected within the present approximation scheme. It also vanishes for 
$\alpha = \frac{1}{2}$. Since we will later find that the effective action depends only on $\bar{\omega}$ and is thus independent of $\alpha$ we can safely put $\alpha= \frac{1}{2}$. The remaining
derivative acting on eq.~(\ref{3.21}) gives the gap equation in the form
\begin{align}
0 = \frac{\delta F}{\delta \wb(k)} = 
 - \frac{V_d}{(2\pi)^d}\,&\frac{1}{2 \wb(k)^2}\cdot \Bigg\{ (N^2 -1)\,d_\xi\,
\big[\,k^2 - \wb(k)\,\big] +
\nonumber \\[1mm]
&+ g^2\,C_2 \,\int \frac{d^dp}{(2\pi)^d}\,\left[ b_1 - b_2 \frac{(k\cdot p)^2}{k^2 p^2}\right]
\frac{1}{\wb(p)} \Bigg\} - \frac{\delta}{\delta \wb(k)}\,\langle \ln \mathscr{J}\rangle_{\wb}\,. 
\label{3.23}
\end{align}
For the last term, we resort again to the curvature approximation eq.~(\ref{3.10}).
With the correlators from eq.~(\ref{3.13.1}), we have
\begin{align}
\langle \ln \mathscr{J}\rangle_{\wb} &\approx - \frac{1}{2}\int d(x,y)\,\chi^{ab}_{\mu\nu}(x,y)\,
\langle A_\mu^a(x)\,A_\nu^b(y)\rangle_{\wb,\Ac} \nonumber \\[1mm]
&= - \frac{1}{2}\int d(x,y)\,\chi^{ab}_{\mu\nu}(x,y)\,\left [ \Ac^a_\mu(x)\,\Ac_\nu^b(y) + 
\big[\wb^{-1}\big]^{ab}_{\mu\nu}(x,y) \right]
\nonumber \\[2mm]
& \stackrel{\mathcal{A} = 0}{=}  
- \frac{1}{2}V_d \,d_\xi\, (N^2 -1)\,\int\frac{d^d p}{(2\pi)^d}\,\chi(k)\,\wb(k)^{-1} 
\,.\label{3.24}
\end{align}
Taking the derivative w.r.t.~$\wb(k)$ and neglecting again the implicit dependence
of $\chi$ on $\wb$, we obtain
\begin{align}
 - \frac{\delta}{\delta \wb(k)}\,\langle \ln \mathscr{J}\rangle_{\wb}
\approx - \frac{1}{2}\frac{V_d}{(2\pi)^d}\,d_\xi\, (N^2 -1) \cdot \frac{\chi(k)}{\wb(k)^2}\,.
\label{3.25}
\end{align}
We can now use this result in eq.~(\ref{3.23}) and finally obtain the gap equation in 
the form
\begin{align}
 \wb(k) = k^2 + \chi(k) + \frac{N g^2}{d_\xi}\,\int \frac{d^d p }{(2\pi)^d}\,
&\Big[ b_1 - b_2 \,\frac{(k\cdot p)^2}{k^2 p^2}\Big]\,\frac{1}{\wb(p)} 
\,. 
\label{3.30}
\end{align}
When the solution of eq.~(\ref{3.30}) is used in the traditional way to compute 
Schwinger functions and eventually the effective action, the result \emph{must} be 
\begin{align}
\Gamma(\Ac) = \frac{1}{2}\int d(x,y)\Ac^a_\mu(x)\,\Ac^b_\nu(y)\cdot
\wb^{ab}_{\mu\nu}(x,y)\,, 
\label{3.32}
\end{align}
because the gluon propagator at $\Ac = 0$ is $\wb^{-1}$ according to eq.~(\ref{3.13.1}),
and the functional approach has no higher vertices.  

Due to the Lorentz invariance of our approach at $\Ac = 0$, the kernel $\wb(p)$ is a 
function of $|p|$ only and does not single out a direction, so  that
\[
\int \frac{d^d p}{(2\pi)^d}\,\frac{p_\mu p_\nu}{p^2}\,\frac{1}{\wb(p)} = 
\frac{\delta_{\mu\nu}}{d} \int \frac{d^d p}{(2\pi)^d}\,\frac{1}{\wb(p)}\,.
\]
The gap equation (\ref{3.30}) can now be written in a very compact form,
\begin{align}
\wb(k) = k^2 + M^2 + \chi(k)\,, 
\label{3.50}
\end{align}
where the gluon mass is dynamically generated through the non-linear integral equation,
\begin{align}
M^2 = C\cdot N g^2\,\int \frac{d^d p}{(2\pi)^d}\,\frac{1}{p^2 + M^2 + \chi(p)}\,,
\label{3.52} 
\end{align}
with $C \equiv (b_1 - b_2/d)/d_\xi$. The gap equation (\ref{3.50}) is very 
transparent: the dynamical mass $M^2$ is generated from the 4-gluon vertex, 
while the curvature $\chi(k)$ describes the coupling to the Faddeev-Popov ghost fields. 
As we will see shortly $\chi(k)$ is just the ghost loop, see eq.~(\ref{c7}) below.
 In fact the gap equation can be interpreted as the dispersion relation of a 
relativistic particle with mass $M$ and a self-energy given by the curvature $\chi(k)$.

\medskip\noindent
Let us finally emphasize that the functional approach and the gap equation (\ref{3.50}) can be 
interpreted in a slightly different way: If we go back to the general definition eq.~(\ref{1.5}) 
of the effective action, but this time constrain the gluon propagator 
$\langle A^a_\mu(-p)\,A^b_\nu(p)\rangle = \delta^{ab} t_{\mu\nu}(p)\,\wb(p)^{-1}$ 
instead of the gluon field itself, we can set $\Ac = 0$ in our trial measure eq.~(\ref{3.13}) and 
interprete the variational parameter $\wb$ as the \emph{classical value} for the inverse gluon 
propagator. The free action $F(\wb)$ from eq.~(\ref{3.21}) therefore coincides with the 
effective action $\Gamma(\wb)$ for the (inverse) gluon propagator. This entails that the 
optimal value for $\wb$ is given by $\delta \Gamma / \delta \wb = \delta F(\wb)/\delta \wb  = 0$, 
which is exactly eq.~(\ref{3.30}) and hence the gap eq.~(\ref{3.50}). 
\emph{Thus, the gap equation yields the best match with the exact gluon propagator (in the sense of 
the effective action) which can be achieved within our variational ansatz.}
No such argument exists for the vertex corrections in the linear response approach, i.e.~while
this formulation is able to produce radiative corrections for higher-order Green functions 
(even with a Gaussian measure), a realistic description of higher vertices presumably requires 
to go beyond the Gaussian \emph{ansatz} \cite{Campagnari:2010wc}. 
We will therefore restrict our investigations to the propagators of the theory.
 
\subsection{Ghost DSE and the curvature}
\label{app:chi}

The gap equation (\ref{3.50}) contains the curvature $\chi(k)$ (\ref{3.9}), which is nothing but the ghost loop and will be calculated below.
The ghost propagator is the expectation value (in our trial measure) of the inverse 
Faddeev-Popov operator
\begin{align}
G^{-1} = - \hat{D}_\mu\,\partial_\mu = - (\partial_\mu + g \hat{A}_\mu)\,\partial_\mu \equiv G_0^{-1} - h\,, 
\label{c1}
\end{align}
where $G_0 = - \Box$ is the free ghost propgator and $h = g\hat{A}\partial$ describes the 
interaction with the gluon.\footnote{Unless stated otherwise, all operators in this subsection 
are adjoint colour and spacetime matrices, for instance $G = G^{ab}(x,y)$.}
From the usual resolvent identities, we obtain first $G = G_0 + G_0\, h\, G$ and thus
\begin{align}
\langle G \rangle = G_0 + G_0 \,\langle h\,G\rangle =: G_0 + G_0\,\Sigma\,\langle G \rangle\,,
\label{c2}
\end{align}
where we have introduced the ghost self energy $- \Sigma = \langle G \rangle^{-1} - G_0^{-1}$. 
Following Ref.~\cite{Feuchter:2004mk} one can derive for $\Sigma$ the expression
\begin{align}
\Sigma \,\langle G \rangle \equiv \langle h\,G\rangle = \int D\,\Gamma_0\,\langle G \rangle \,\Gamma\,,
\label{c3} 
\end{align}
which involves the gluon propagator $D$ from eq.~(\ref{G10}) as well as the free ($\Gamma_0$) 
and full ghost-gluon vertex $\Gamma$ defined by 
\begin{align}
\langle\,G\,\Gamma_0\,G\,\rangle = \langle\,G\,\rangle \,\Gamma\,\langle\,G\,\rangle\,.
\label{G11} 
\end{align}
From Dyson-Schwinger and flow equation approaches to Yang-Mills theory in  Landau gauge, 
it is well known that the dressing of the full ghost-gluon vertex is a subleading effect, 
so that the full vertex in the last equation can be replaced by the bare one 
(\emph{rainbow-ladder approximation}). This has the advantage that 
the ghost self-energy $\Sigma$ (and thus the full ghost propagator $G$) can be expressed through 
the kernel $\wb$ alone. To do so, we introduce the \emph{ghost form factor} $\eta$ via
\begin{align}
\langle G \rangle = G_0 \cdot \eta\,.
\label{c4} 
\end{align}
Note that both $G_0$ and $\langle G \rangle$ are colour diagonal, so that $\eta = \eta(x,y)$ 
has no colour index. Using eq.~(\ref{c2}), it is easy to see that $\eta^{-1} = 1 - \Sigma\,G_0$.
If we now use the rainbow-ladder approximation in eq.~(\ref{c3}) for $\Sigma$ and re-express 
$\langle G \rangle$ through the form factor via eq.~(\ref{c4}), we obtain a closed integral 
equation for the ghost form factor in momentum space,
\begin{align}
\eta(k)^{-1} = 1 - N g^2\,I_\eta(k) \equiv 1 - 
Ng^2\,\int \frac{d^d q}{(2\pi)^d}\,\Big[ 1 - (\hat{k}\cdot \hat{q})^2 \Big]\,
\frac{\eta(k-q)}{(k-q)^2\,\wb(q)}\,,
\label{3.54}
\end{align}
where the explicit form eq.~(\ref{G10}) of the gluon propagator was inserted.

As for the curvature eq.~(\ref{3.9}), taking two functional derivatives of 
$(\ln \mathscr{J}) = \mathrm{Tr}\ln G^{-1}$ and using eq.~(\ref{G11}) yields
\begin{align}
\chi^{ab}_{\mu\nu}(x,y) = - \mathrm{Tr}\,\Big[ \langle G \rangle \,\Gamma^a_\mu(x)\,\langle G 
\rangle\,[\Gamma_0]^b_\nu(y)\Big]\,.
\label{c7} 
\end{align}
Using the rainbow-ladder approximation again and contracting Lorentz indices and colours, we can 
express the scalar curvature eq.~(\ref{3.11}) in momentum space through the ghost form factor, 
\begin{align}
\chi(k) = N g^2\,I_\chi(k) \equiv  Ng^2 \cdot \frac{1}{d-1}\,\int \frac{d^d q}{(2\pi)^d}\,
\Big[1 - (\hat{k}\cdot \hat{q})^2\Big] \,\frac{\eta(k-q)\,\eta(q)}{(k-q)^2}
\label{3.54.1}
\end{align}
The DSE (\ref{3.54}) for the ghost form factor depends explicitly on the kernel $\wb$ of the trial 
measure (\ref{3.13}), determinated by the gap equation (\ref{3.50}).
Eqs.~(\ref{3.54.1}), (\ref{3.54}), (\ref{3.52}) and (\ref{3.50}) form a closed system to 
determine the ghost form factor $\eta(k)$, the curvature $\chi(k)$, the mass $M$,
 and the variational kernel $\bar{\omega}(k)$. 
 
%

\section{Renormalization}
\label{sec:5}

\subsection{Counterterms}

\label{sec::ren}
To complete our analysis, we have to determine the high momentum behaviour of our 
Green functions and renormalise the corresponding integral equations. 
We begin with the ghost DSE (\ref{3.54}). For large momenta, $\wb(k) \sim k^2$ and $\eta(k) \sim 1$ 
(up to logarithmic corrections), so that dimensional analysis implies for the 
logarithmic divergence in $I_\eta(k)$ (\ref{3.54})
\begin{align*}
\eta^{-1}(k) = 1 - \Sigma\,G_0 = 1 - Ng^2\,I_\eta(k)= 1 - Ng^2 \,\Big[a_0\,\ln (\Lambda^2/M^2) + 
\text{finite}\Big]\,,
\end{align*}
where $\Lambda$ is a suitable UV cutoff and $a_0$ a finite numerical constant that depends on
the cutoff procedure. Since $G_0(k) = 1/k^2$, the corresponding counterterm $\delta \Sigma$ for 
the ghost self-energy is proportional to $k^2$, i.e.~it is a \emph{ghost field renormalisation}. 
(No ghost mass term is induced by the theory.) The same conclusion could be reached if we introduced 
explicit ghost fields $\{c,\bar{c}\}$, because $\langle \bar{c}\,c\rangle = \langle G \rangle = G_0\,\eta$ 
and the field renormalisation $c \to \sqrt{Z_c}\,c$ is equivalent to $\eta \to Z_c\,\eta$. 
To the given loop order, this field renormalisation leads to 
\begin{align*}
\frac{1}{\eta(k)} = 1 - N g^2\,I_\eta(k) - \delta Z_c + \text{two loops}\,. 
\end{align*}
In terms of explicit ghost fields, the counterterm $\delta Z_c = (\delta \Sigma) G_0$ 
(or $\delta \Sigma = k^2\,\delta Z_c$) would hence correspond to the local expression 
\begin{align}
\delta Z_c\,  \int d^4x\,\partial_\mu \bar{c}\,\partial^\mu c 
\end{align}
in the exponent of the trial measure (\ref{3.13}).
We can now adjust the finite pieces in $\delta Z_c$ such that
\begin{align} 
1 - \delta Z_c = \eta(\mu)^{-1} + Ng^2\,I_\eta(\mu)
\label{3.500}
\end{align}
where $\eta(\mu)$ is an arbitrary finite constant, because eq.~(\ref{3.500}) is independent of $k$ 
and the (loga\-rithmic) $\Lambda$-divergences on both sides agree.  This prescription leads 
directly to the renormalised ghost equation 
\begin{align}
\eta(k)^{-1} = \eta(\mu)^{-1} - Ng^2\,\Big[ I_\eta(k) - I_\eta(\mu) \Big]\,, 
\label{3.55.1}
\end{align}
which could also be obtained by simply subtracting the bare equation (\ref{3.54}) at 
$k=\mu$. Notice that eq.~(\ref{3.500}) may be a rather unusual field normalisation, but any other 
prescription for $Z_c$ can only differ by a finite constant. Notice also that the renormalised 
eq.~(\ref{3.55.1}) is independent of $g$, as can be seen e.g.~by rescaling 
$\eta \to \widetilde{\eta} \equiv g\,\eta$.  

\bigskip\noindent
Next, we study the mass and curvature equations (\ref{3.52}) and  (\ref{3.54.1}), respectively, 
which are quadratically divergent by power counting,
\begin{align}
M^2 &= Ng^2\,\Big[ a_1\,\Lambda^2 + b_1 \,M^2\,\ln(\Lambda^2 / M^2) + \text{finite}\Big] 
\nonumber\\[2mm]
\chi(k) &= Ng^2 \,\Big[ a_2 \,\Lambda^2 + b_2\,M^2\,\ln(\Lambda^2/M^2) + 
c\,k^2\,\ln(\Lambda^2/M^2) + \text{finite}\Big]\,,
\label{3.1024}
\end{align}
with numerical factors $a_i$, $b_i$ and $c$ that depend on the cutoff procedure.\footnote{Eq.~(\ref{3.1024})
is consistent with Ref.~\cite{Reinhardt:2010ar} where the divergencies of the Faddeev-Popov determinant 
were identified within a gradient expansion.}
The subtraction is a bit more complicated in this case because of the sub-leading 
logarithmic divergence. We begin by subtracting the $k$-independent contributions with 
counterterms for $M^2$ and $\chi(k)$, respectively,
\begin{align}
\delta M_1^2 &\equiv - Ng^2\,\Big[ a_1\,\Lambda^2 + b_1\,M^2 \,\ln(\Lambda^2 / M^2)
+ \text{finite} \Big] 
\nonumber \\[2mm]
\delta\chi_1 &\equiv - N g^2 I_\chi(\mu) + \chf\,,
\label{3.1025}
\end{align}
where $\chf$ is an arbitrary finite constant. 
The mass equation (\ref{3.52}) is now finite with a dynamically induced mass $M^2$ whose absolute value 
is undetermined because of the finite pieces in the corresponding counterterm $\delta M_1^2$.
As a consequence, we do not need to solve eq.~(\ref{3.52}) but rather take $M^2$ as a finite 
parameter that can be chosen at will. The subtracted curvature equation takes the form 
\begin{align}
\chi(k) = N g^2\,\Big[ I_\chi(k) - I_\chi(\mu)\Big] + \chf \,. 
\label{3.1026}
\end{align}
This is not yet finite because the difference of the two integrals contains the subleading 
logarithmic divergence (cf.~eq.~(\ref{3.1024})) 
\begin{align}
\Big[ I_\chi(k) - I_\chi(\mu)\Big] = c \,(k^2-\mu^2)\,\ln(\Lambda^2 / \mu^2) + \text{finite.}
\label{3.1028}
\end{align}
We must therefore add a second, $k$-dependent counterterm $\delta\chi_2(k)$ which equals the negative
of the divergence on the rhs of eq.~(\ref{3.1028}). In order to associate these subtractions 
with local terms in the exponent of the trial measure eq.~(\ref{3.13}), we isolate the 
$k$-dependent pieces in $\delta \chi_2(k)$ and write the total curvature counterterm as 
$\delta \chi(k) \equiv \delta \chi_1 + \delta \chi_2(k) = \delta \chi_0 + k^2\,\delta Z_A$ with
\begin{align}
\delta \chi_0 &\equiv - N g^2 \,I_\chi(\mu) + c\,\mu^2\,\ln(\Lambda^2/\mu^2) + \chf
\nonumber \\[2mm]
\delta Z_A &\equiv - N g^2\,c\,\ln(\Lambda^2/\mu^2) + z(\mu)\,,
\end{align}
where $z(\mu)$ is again an arbitrary finite piece that can be added to the counterterm.
If we now replace $\chi(k) \to \chi(k) + \delta \chi(k)$ as well as  
$M^2 \to M^2 + \delta M_1^2$, eqs.~(\ref{3.52}) and (\ref{3.54.1}) will be finite.
From the gap equation (\ref{3.50}), it is clear that the $k$-independent subtractions 
can be combined to a \emph{gluon mass counterterm} $\delta M^2 = \delta \chi_0 + \delta M_1^2$, 
while the $k$-dependent counter term $k^2\,\delta Z_A$ rescales the kinetic 
energy, i.e.~it represents a gluon field renormalisation. These subtractions 
correspond to local counterterms
\begin{align}
\frac{1}{2}\,\delta M^2\, \int d^4x\,(A_\mu^a)^2
+ \frac{1}{4}\,\delta Z_A\,\int d^4x\, (\partial_\mu A_\nu^a - \partial_\nu A_\mu^a)^2
\label{3.224}
\end{align}
in the exponent of the trial measure (\ref{3.13}) (with $\mathcal{A} = 0$). 
The renormalised curvature equation is now
\begin{align}
\chi(k) = N g^2\,\Bigg[ I_\chi(k) - I_\chi(\mu) - c\,(k^2 - \mu^2)\,\ln(\Lambda^2/\mu^2) \Bigg]
+ z(\mu)\,(k^2-\mu^2) + \chi(\mu)\,, 
\label{3.442}
\end{align}
where $\chi(\mu) = \chf + z(\mu)\,\mu^2$ is finite.

\medskip
Eq.~(\ref{3.442}) is not very suitable for numerical investigations. To put it in a manageable form, 
we need to determine the constant $c$ and devise a way to perform the necessary subtractions 
under the integral $I_\chi(k)$. Unfortunately, $I_\chi(k)$ and hence the factor $c$ depend on the 
ghost profile $\eta(k)$ and its derivatives, which are only known numerically. Thus, we proceed as follows: 
At any stage in the iterative solution of the integral equation system, we perform the angular integration 
in $I_\chi(k)$ with the current solution of the ghost form factor $\eta(q)$ to find an expression
\begin{align}
I_\chi(k) = \int\limits_0^\Lambda\,dq\,f(k,q)\,,
\label{3.4500}
\end{align}
with a complicated function $f$ only known numerically. Our construction of the counterterms above
translates into the asymptotics
\begin{align}
f(k,q) - f(\mu,q) \stackrel{q \to \infty}{\longrightarrow}\,\frac{(k^2 - \mu^2)}{q}\cdot c + \mathscr{O}(q^{-2})\,. 
\label{3.1060}
\end{align}

\renewcommand{\arraystretch}{1.4} 
\begin{table}[t!]
\centering
\begin{tabular}{ccc}
\toprule
renormalisation & counter term & renormalisation constant  \\ \colrule
gluon field & $\delta Z_A$  & $z(\mu)$   \\
gluon mass  & $\delta M^2 = \delta \chi_0 + \delta M_1^2$ & $M_A^2(\mu)$ \\
ghost field &  $\delta Z_c$ & $\eta(\mu)$\\ \botrule
\end{tabular}
\caption{\label{tabx}Renormalisation of the final system of integral equations.}
\end{table}
\renewcommand{\arraystretch}{1.0}

\begin{figure}[t]
 \centering
 \includegraphics[width=7.5cm]{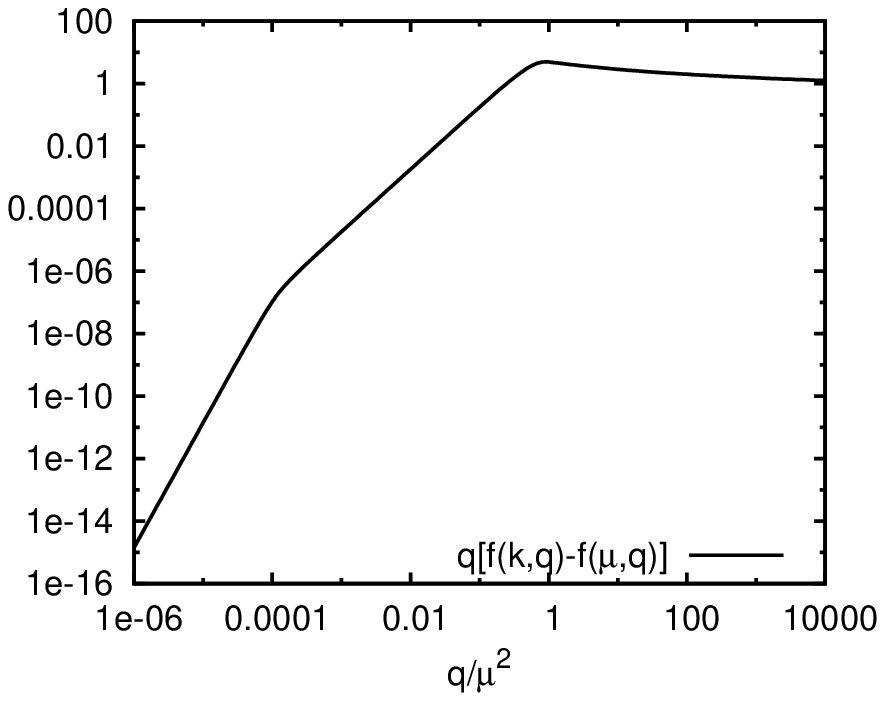}
\hspace*{\fill}
 \includegraphics[width=7.5cm]{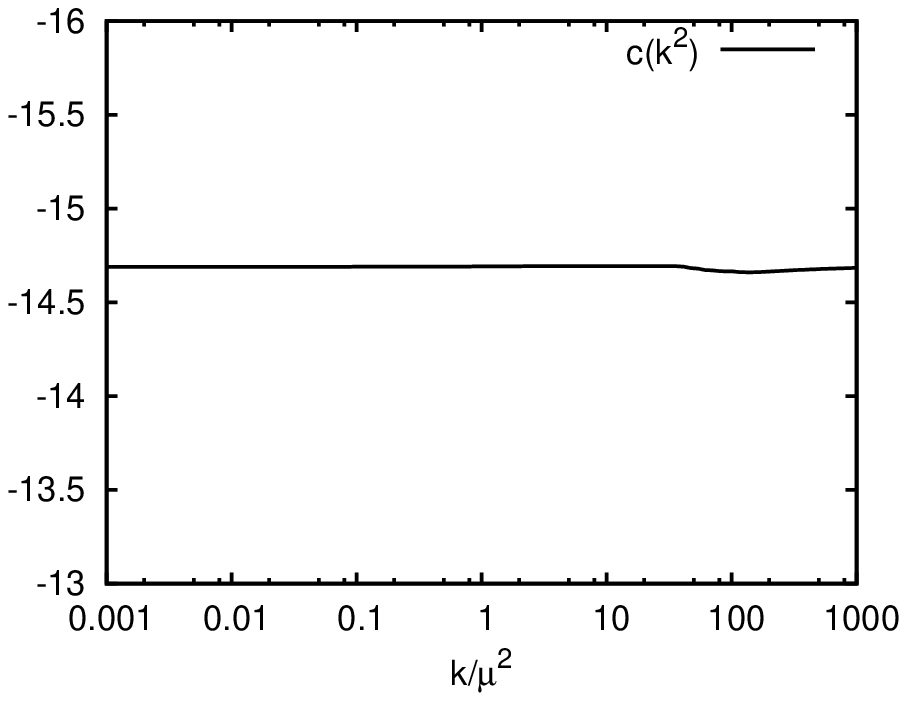}
 \caption{\emph{Left panel:} The scaled integrand $q\big[ f(k,q)-f(\mu,q)\big]$ of the 
subtracted loop integral $I_\chi(k)$ as a function of $q$, cf.~eq.~(\ref{3.1060}).
\emph{Right panel:} The asymptotic value of the subtracted integrand in $I_\chi(k)$, 
as a function of $k$, cf.~eq.~(\ref{3.1064}).}
 \label{fig:x}
\end{figure}

This can be verified numerically: In the left panel of fig.~\ref{fig:x}, we have plotted 
$q\big[ f(k,q)-f(\mu,q)\big]$ as a function of $q$ and observe that it approaches a constant 
value\footnote{There are still small deviations from a constant due to numerical
issues that prevent us from going to very large cutoffs while preserving sufficient accuracy in 
the iterative solution of the integral equation system. However, extracting $c$ as in eq.~(\ref{3.1064}) 
yields finite iterative solutions that become cutoff-independent long before the numerics become delicate.} 
for large $q$. The value of that constant depends on $k$ in the expected way: In the right panel of 
fig.~\ref{fig:x}, we have plotted the value 
\begin{align}
\frac{q}{k^2 - \mu^2}\,\big[ f(k,q) - f(\mu,q)\big]\Big\vert_{q = \Lambda} 
\label{3.1064}
\end{align}
as a function of $k$ and observe that this quantity is independent of $k$, as expected 
from eq.~(\ref{3.1060}). The constant value is exactly the factor $c$ used in eq.~(\ref{3.442}) above. 
Thus, at any stage of the iterative solution with the current form of $\eta(k)$, we first perform 
the angle integrations in $I_\chi(k)$ to obtain the integrand $f(k,q)$ in eq.~(\ref{3.4500}) and 
then evaluate the curvature $\chi(k)$ from eq.~(\ref{3.442}) written in the form
\begin{align}
\chi(k) &= N g^2 \,\int\limits_0^\infty dq\,\Bigg[ f(k,q) - f(\mu,q) - 
\frac{\Lambda}{q}\,\big[f(k,\Lambda)-f(\mu,\Lambda)\big]\Bigg] 
+ z(\mu)\,(k^2-\mu^2) + \chi(\mu)\,.
\label{3.1066} 
\end{align}
This procedure determines the counterterm coefficient $c$ from eq.~(\ref{3.442}) iteratively
and renders the system of integral eqs.~(\ref{3.50}), (\ref{3.55.1}) and (\ref{3.1066}) finite. 
It also leads to a stable solution which is numerically independent of the cutoff 
$\Lambda$ if the latter is chosen large enough. The system involves four renormalisation 
constants $\eta(\mu)$, $\chi(\mu)$, $z(\mu)$ and $M^2$ which could even be defined at different 
scales $\mu$. One of the constants is, however, redundant because finite changes in $\chi(\mu)$ can 
be absorbed in $M^2$ as discussed above. In fact, only the \emph{gluon mass parameter}\footnote{To avoid problems with possible infrared 
singularities, we renormalize the gluon propagator at $\mu > 0$ so that 
$M_A^2(\mu) = M^2 + \chi(\mu) - z(\mu)\,\mu^2$ is merely a renormalization constant without a direct 
interpretation as a mass. If the curvature (and thus the gluon propagator) happens to be finite at $k=0$, 
the intercept $\wb(0) = M^2 + \chi(0) = M_A^2(0)$ can be interpreted as a (constitutent) gluon mass.} 
$M_A^2(\mu) \equiv M^2 + \chi(\mu) - z(\mu)\,\mu^2$ will appear in the renormalized equations, and we can 
therefore drop either $\chi(\mu)$ or $M^2$ in favour of the other without loss of generality. 
The remaining renormalisation constants $\eta(\mu)$ and $z(\mu)$ fix the prefactor of the kinetic 
energy for ghost and gluon, respectively, i.e.~they fix the scale of the ghost and gluon field.  
(No vertex renormalisation is induced by the theory.) The three independent counter terms are also summarised in table \ref{tabx}. 
 
\subsection{Numerical treatment}
Let us briefly comment on the numerical treatment of the integral equation system, 
and in particular on the role of the various renormalization constants. First, we note that
the coupling constant $g$ can be eliminated from all equations by rescaling 
$\tilde{\eta} = g\,\eta$. 
If we insert the curvature equation (\ref{3.1066}) directly into the gap equation (\ref{3.50}) 
and further introduce the gluon mass parameter $M_A^2(\mu) \equiv M^2 + \chi(\mu) - z(\mu)\,\mu^2$ as 
before, we obtain the renormalized system
\begin{align}
\wb(k) &= \big[1+z(\mu)\big]\,k^2 + M_A^2(\mu) + 
N\,\Big[I_\chi(k) - I_\chi(\mu) - c (k^2 - \mu^2)\,\ln(\Lambda^2/\mu^2)
\Big] 
\nonumber \\[2mm]
\tilde{\eta}(k)^{-1} &= \tilde{\eta}(0)^{-1} - N \,\Big[ I_{\tilde{\eta}}(k) - I_{\tilde{\eta}}(0)\Big]\,.
\label{R1} 
\end{align}
Recall that the coefficient $c$ in the first equation is determined during the iterative solution
of the system such that it becomes cutoff-independent. 

\medskip
Ideally, we would choose the renormalization point $\mu$ for the gluon and ghost field 
at a large Euclidean scale $\mu \gg 1$ far away from possible singularities, where 
asymptotic freedom provides natural renormalization conditions. Unfortunately, 
such a procedure does not lead to a stable numerical solution in the deep infrared, 
because the system has a family of solutions which are qualitatively different at low momenta,
but cannot be discriminated in the ultraviolett. In order to stabilize the integration and 
bring the distinct solutions to the fore we must, at least for the ghost equation, choose 
a renormalization point in the deep infrared, or even at $\mu=0$, which we did in all numerical 
investigations. The gluon renormalization scale $\mu$, by contrast, can be chosen finite and arbitrary, 
and we use it to rationalize all dimensionfull quantities within our 
numerical treatment.
Thus, we find  that the shape of the rescaled (dimensionless) integral equation 
system only depends on the three dimensionless parameters $\tilde{\eta}(0)^{-1}$, $z(\mu)$, and 
$M_A(\mu)/\mu$. 


We have already mentioned above that the first of these three parameters, $\tilde{\eta}(0)^{-1}$, 
discriminates between the scaling and decoupling solutions. As for the gluon mass parameter $M_A(\mu)/\mu$, 
it has no effect on the infrared behaviour of the scaling solution, while it determines 
the $k=0$ limit of the gluon propagator for the decoupling solution, i.e.~the constitutent gluon mass.
Both renormalisation parameters have negligable effect on the deep infrared behaviour but determine 
the momentum scale at which the transition to the infrared behaviour sets in.

Finally, the gluon  field renormalization factor $z(\mu)$ changes the overall size of the 
kernel $\wb(k)$. More precisely, any change $z(\mu) \to z'(\mu)$ for the scaling type of solution 
simply leads to an overall rescaling
\begin{align}
\wb'(k) = \frac{1+z'}{1+z}\,\wb(k)\,,\qquad\quad 
\tilde{\eta}'(k) = \sqrt\frac{1 + z'}{1+z}\,\tilde{\eta}(k) \,.
\label{R2}
\end{align}
For the decoupling solution, this simple rescaling also holds if we simultaneously 
change the finite renormalization constant $\tilde{\eta}^{-1}(0)$ as in eq.~(\ref{R2}).

\section{Results}
\label{sec:6}

\subsection{Infrared analysis}
\label{sec::ir}
The infrared analysis of the renormalized system of integral equations (\ref{R1}) can be carried out much 
as in the Coulomb gauge case \cite{Schleifenbaum:2006bq,Heffner:2012sx}, see also 
Ref.~\cite{Lerche:2002ep} and \cite{Zwanziger:2003de}. If we assume a power-like behavior in the infrared,
\begin{align}
\wb(k) \sim (k^2)^{\alpha}\,,\qquad\qquad \eta(k) \sim (k^2)^{-\beta}  
\label{3.56}
\end{align}
the non-renormalisation of the ghost-gluon vertex \cite{Taylor:1971ff} implies the \emph{sum rule}
\begin{align}
- \alpha + 2 \beta = \frac{d}{2} - 1\,.
\label{3.60} 
\end{align}
Furthermore, if $\bar{\omega}$ is infrared divergent it follows from the gap equation (\ref{3.50}) that the curvature 
$\chi(k)$ (\ref{3.54.1}) also has the same infrared exponent $\alpha$. 
Depending on the choice of our finite renormalization constants we obtain two different types of solutions:
\begin{itemize}
 \item[\textbf{(i)}] { Implementing the horizon condition \cite{Zwanziger:1991ac} $\eta^{-1}(0) =0$ we find the so-called critical or scaling solution with the infrared exponents
    \begin{equation}
    \beta = \frac{1}{98}\left( 93 \mp \sqrt{1201})\right) \approx 
     \big\{\,0.5954,\,\,\, 1.3025\,\Big\},
    \label{3.60.1}
    \end{equation}
    which both entail $\alpha > 0$, i.e.~an infrared vanishing gluon propagator. These are the same 
    infrared exponents found from the one loop DSEs in Landau gauge \cite{Lerche:2002ep}. Only the first of 
    these scaling solutions could be found in our numerical calculation, cf.~sect.~\ref{sec:num}.
    It should also be mentioned  that the value of the gluon mass parameter $M^2_A(\mu)$ 
    is irrelevant, as long as the curvature $\chi(k)$ is infrared divergent, i.e.~as long as $\beta > 1/2$, 
    which is the case for both exponents given in eq.~(\ref{3.60.1})}.
 \item[\textbf{(ii)}] {
    Assuming an infrared finite ghost form factor $\eta^{-1}(0) >0$, i.e. $\beta =0$, the sum rule 
    (\ref{3.60}) would yield $\alpha = -1$. However, in this case a non-zero mass parameter $M^2_A(\mu)$ 
    in the gap equation dominates the infrared behaviour, which invalidates the sum rule. 
    For fixed mass parameter $M_A(\mu)/\mu$ and fixed gluon wave function renormalization $z(\mu)$, 
    we obtain a one-parameter family of solutions labeled by the ghost renormalization constant   
    $\eta^{-1}(0) \neq 0$. These are the so-called subcritical or decoupling solutions already found 
    from the DSEs in Landau gauge \cite{Lerche:2002ep}.} The intercept $\wb(0) = M^2 + \chi(0)$ can be 
    interpreted as a constituent gluon mass.
\end{itemize}
In the special case $M_A^2(\mu) =0$ there is also a solution with an infrared finite ghost form factor
$\eta(k)$ and an infrared vanishing gluon kernel $\wb(k)$, which formally obeys the sum rule 
with $\alpha=-1$ and $\beta=0$. This solution is, however, definitely ruled out by all existing 
lattice data, in which the ghost form factor always diverges at $k \to 0$. This leaves us with 
just the two type of solutions listed above. In the next subsection we compare these solution 
with high-precision lattice results.

\subsection{Numerical results}
\label{sec:num}
As discussed above the discriminating criterion for the two type of solutions is whether we 
choose the ghost form factor $\eta(0)$ finite (\emph{decoupling solutions}) or 
infinite (\emph{scaling solution}). Let us briefly describe our numerical findings for both 
type of solutions:

\begin{figure}[t]
 \centering
 \includegraphics[width=7.5cm]{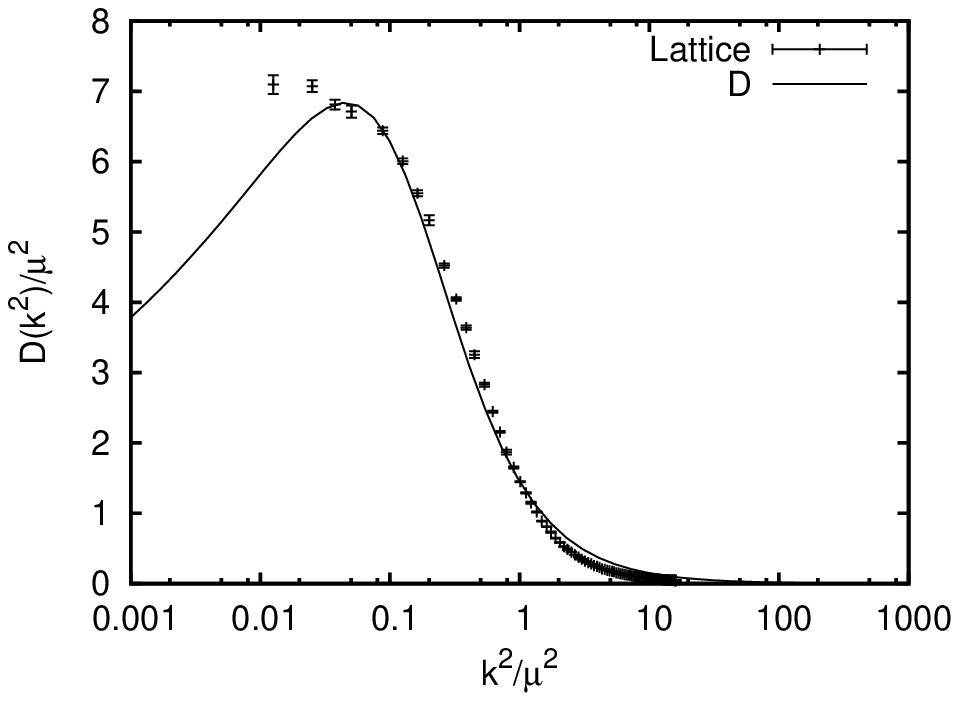}
\hspace*{\fill}
 \includegraphics[width=7.5cm]{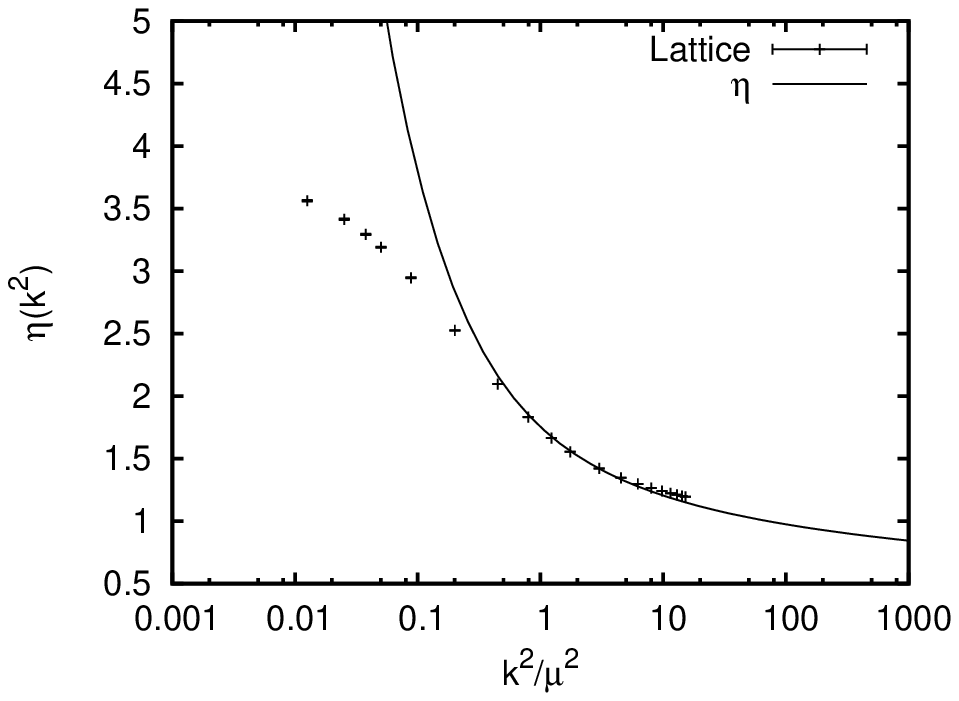}
 \caption{Critical solution for the gluon propagator (left panel) and the ghost form factor 
(right panel) for SU$(2)$ Yang-Mills theory in Landau gauge. 
The crosses denote the lattice results from Ref.~\cite{Bogolubsky:2009dc}.}
 \label{fig:x1}
\end{figure}

\medskip\noindent 
\paragraph*{\textbf{Scaling solution:}} With $\eta(0)^{-1} = 0$  we have \emph{enforced} an 
IR diverging ghost form factor. In our numerical solution we could only find the less divergent 
of the two possible exponents determined in the IR analysis eq.~(\ref{3.60.1}). The best fit to
our numerical data suggests $\beta = 0.595(3)$ for the ghost form factor and $\alpha = 0.191$ 
for the gluon kernel $\wb$, which are both in excellent agreement  with the analytic results 
eq.~(\ref{3.60.1}). Furthermore the sum rule eq.~(\ref{3.60}) is satisfied numerically to better than  
$10^{-3}$ accuracy. To compare with lattice data, we have to adjust the finite gluon wave function 
renormalization $z(\mu)$ at an arbitrary scale $\mu > 0$ to match the overall scale of the propagator.
The remaining renormalization parameter $M_A^2(\mu)$ is immaterial since it has no effect on 
the final solution. (This is because the IR diverging curvature dominates all constant terms 
in the gap equation at $k \to 0$.) In figure \ref{fig:x1}, the critical solution is compared to 
high-precision lattice data taken from Ref.~\cite{Bogolubsky:2009dc}. It is obvious that the 
scaling type of solution describes the UV behaviour fairly well, but it severly deviates in the 
deep infrared.
  
\begin{figure}[t]
 \centering
 \includegraphics[width=7.5cm]{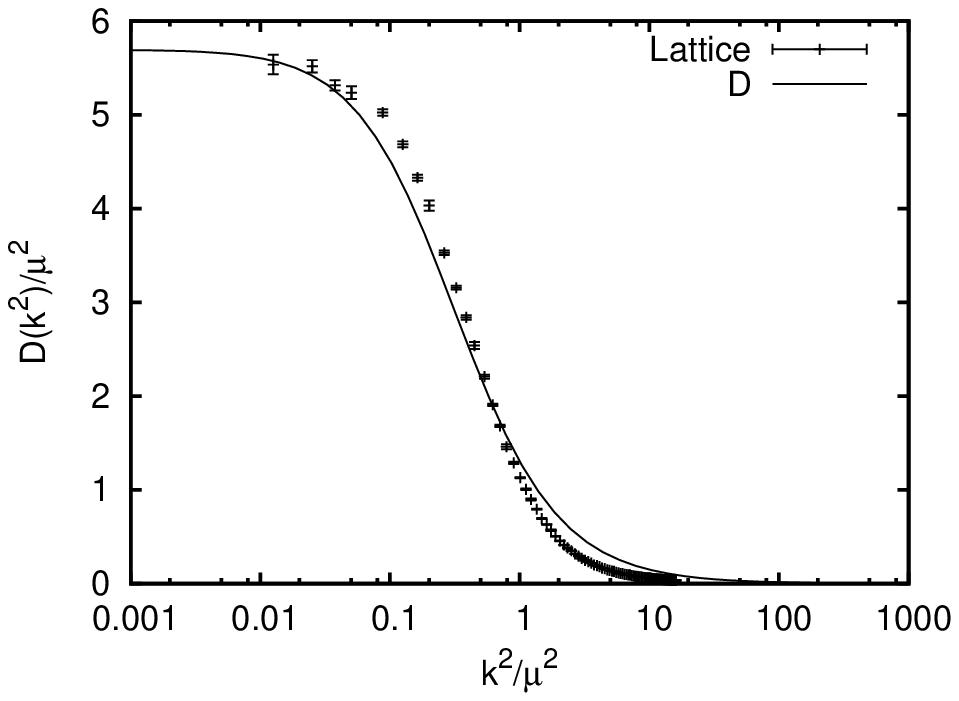}
\hspace*{\fill}
 \includegraphics[width=7.5cm]{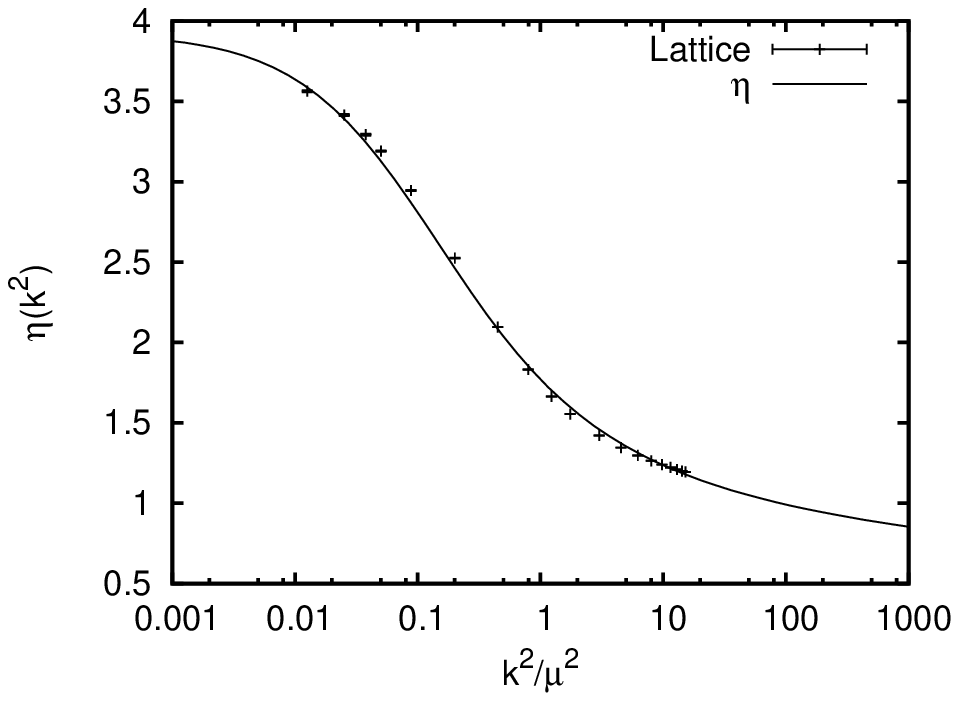}
 \caption{Subcritical solution for the gluon propagator (left panel) and the ghost form factor 
(right panel) for $SU(2)$ Yang-Mills theory in Landau gauge. 
The crosses denote the lattice results from Ref.~\cite{Bogolubsky:2009dc}.}
 \label{fig:x2}
\end{figure}

\bigskip\noindent
\paragraph*{\textbf{Decoupling solutions:}} These solutions have no scaling behaviour, and 
instead exhibit the emergence of a soft BRST breaking mass scale in the deep infrared, 
so that all Green's functions remain finite at $k \to 0$. In this case, the renormalization 
constant $M_A^2(\mu)$ obviously matters as it dominates the gluon propagator in the deep infrared.
In addition, we can also adjust the intercept $\eta(0)$ for the ghost form factor, and the 
overall scale $z(\mu)$ of the gluon propagator at some non-zero scale $\mu > 0$. (The combination of 
$M_A^2(\mu)$ and $z(\mu)$ determines the slope of the gluon propagator in the transition region 
around $k=\mu$.) The decoupling solution is shown in figure \ref{fig:x2}, along with lattice data 
\cite{Bogolubsky:2009dc} for comparison.  As can be seen from the plot, this type of solution 
is in good agreement with the lattice data for the entire momentum range. In particular, 
the agreement with the ghost data is almost perfect.\footnote{We have optimized the 
renormalization parameters in the ghost sector, which incurrs slight deviations for the 
gluon propagator in the transition region $k \approx \mu$, cf.~figure \ref{fig:x2}.
This could be mitigated by a more balanced approach that tries to optimize the 
parameters for both propagators on average.}  

\bigskip\noindent
Unfortunately, the critical and subcritical solution differ only at very low momenta (which is the 
reason for the numerical instability when imposing renormalisation conditions at large scales), 
and it requires large lattices to definitely rule out one or the other. As explained above, 
the available lattice data now clearly favours the decoupling solution. Analytical approaches, by constrast, 
always exhibit both kind of solutions, and it becomes a matter of `boundary conditions' 
$\eta^{-1}(0)$ to select one or the other. In Dyson-Schwinger or functional renormalization group 
approaches, there is no compelling reason to prefer one boundary condition over the other. 
This is different in our variational approach, since it is always the solution with the lowest 
effective action that must be realized. To determine the correct solution, we would thus have to insert 
the various (numerical) solutions $\wb(k)$ in the free action $F$ from eq.~(\ref{3.21}).
This procedure requires first a full renormalization of the effective (or free) action,
which is left for future work.

\section{Summary and conclusions}
\label{sec:x}
In this paper, we have investigated the low-order Green's functions in $SU(N)$ 
Yang-Mills theory, using Landau gauge and a covariant variation principle based on 
the effective action. The formalism leads to a set of integral equations which, 
after proper renormalisation, could be solved numerically over a wide range of 
momenta. We obtain the two types of solutions also found in other functional 
approaches: \textbf{(i)} a critical or scaling solution in which the ghost form 
factor diverges in the infra-red by a power law with an exponent $\beta = 0.5953$ 
while the gluon propagator vanishes with a weak infrared exponent $\alpha = 0.191$, 
and \textbf{(ii)} a  subcritical or decoupling solution where both quantities 
remain finite at low momenta. Recent high-precision lattice data compares favourably with 
both solutions in the UV and into the transition region, 
although detailed studies prefer the subcritical (decoupling) solution in the infrared. 
Our numerical treatment is on par with the best analytical studies, and our decoupling 
solution, in particular, agrees very well with the available lattice data over the 
entire momentum range. In addition, the variation principle used here offers a physical 
criterion to distinguish between the two type of solutions which is not based on 
arbitrary boundary conditions, namely the solution with the lowest free action 
\emph{must} be realized. This question will be studied in a future investigation.

The method presented here works with Euclidean path integrals only, so that it 
naturally generalises to all extensions that can be formulated within a path integral.
In particular, we can include fermions and study finite temperatures and chemical 
potentials without conceptual problems. These issues will also be subject to 
future work.


\begin{acknowledgments}
The authors would like to thank A.~Sternbeck for providing the lattice data used in 
figures \ref{fig:x1} and \ref{fig:x2}. This work was supported in part by DFG under Contract Re-856/6-3 and Re-856/9-1.
\end{acknowledgments}


\begin{appendix}
\section{The quantum effective action}
\label{app:A}
We want to show that the effective action defined by the variation principle
(\ref{1.6}) coincides with the traditional generating functional for 1PI proper
functions. We fix $\varphi(x)$ and start directly from the definition (\ref{1.6}).
First, we have the following upper bound for arbitrary currents $j(x)$,
\begin{align}
\Gamma(\varphi) &\stackrel{(\ref{1.6})}{\equiv}  
\mathop{\mbox{inf}}_\mu \Big[\,\langle\,S\,\rangle_\mu -\hbar\,\As(\mu)  
\, \Big|\,\langle\,\phi\,\rangle_\mu = \varphi\,\Big] \nonumber \\[2mm]
&=\mathop{\mbox{inf}}_\mu \Big[\,\langle\,S\,\rangle_\mu -
\hbar\,\As(\mu) + (j\,,\langle\,\varphi-\phi\,\rangle_\mu) \,
\Big|\,\langle\,\phi\,\rangle_\mu = \varphi\,\Big] \nonumber 
\\[2mm]
& \ge \mathop{\mbox{inf}}_\mu \Big[\,\langle\,S\,\rangle_\mu -
\hbar\,\As(\mu) + (j\,,\langle\,\varphi-\phi\,\rangle_\mu) \Big] \nonumber 
\\[2mm]
&= (j,\varphi) + \mathop{\mbox{inf}}_\mu \Big[\,\langle\,S - (j,\phi)\,\rangle_\mu -
\hbar\,\As(\mu) \,\Big] \nonumber 
\\[2mm]
&\equiv (j,\varphi) - W(j)\,.
\label{X2}
\end{align}
(The inequality follows because the constrained minimum is always larger than the 
unconstrained one.) Next we want to show that $\Gamma(\varphi)$ is in fact the 
\emph{smallest} upper bound, i.e.~the supremum,
\begin{align}
\Gamma(\varphi) = \mathop{\sup}_j \,\Big[\, (j,\varphi) - W(j) \,\Big]\,. 
\label{X4}
\end{align}
To see this, it is sufficient to find a current $j=j_\varphi$ for which the 
bound in eq.~(\ref{X2}) is saturated, 
\begin{align}
\Gamma(\varphi) = (j_\varphi,\varphi)- W(j_\varphi)\,,
\label{X1}
\end{align}
because then the upper bound in eq.~(\ref{X2}) is a maximum 
(and hence a supremum). 
From the derivation of eq.~(\ref{X2}), it is clear that $j_\varphi$ should be chosen 
such as to obey the constraint $\langle \phi \rangle_{\mu_j} = \varphi$ for the Gibbs 
measure $\mu_j$ that solves the minimisation problem on the rhs of eq.~(\ref{X2}).
This is because the unconstrained minimum  happens to obey the constraint 
and is therfore also the constrained minimum, i.e.~the inequality in 
the third line of eq.~(\ref{X2}) becomes an equality. The relevant equation 
$\langle \phi \rangle_{\mu_j} = \varphi$ for $j=j_\varphi$ is, however, merely the 
extremality condition
\begin{align*}
0  = \frac{\delta}{\delta j(x)} \Big[(j,\varphi) - W(j)\Big] = 
\varphi(x) - \frac{\delta W}{\delta j(x)} =  
\varphi(x) -  \hbar\,\frac{\delta Z / \delta j(x)}{Z(j)} = 
\varphi(x) - \langle\,\phi(x)\,\rangle_{\mu_j}\,,
\end{align*}
which we assume to always have a solution.\footnote{This assumption is 
is implicit in the traditional definition of the  generating functional, 
and we do not touch the more subtle question of what happens if no such $j_\varphi$
exists and the supremum (\ref{X4}) is \emph{not} a maximum.}
Thus, the effective action $\Gamma(\varphi)$ defined by the variation principle 
(\ref{1.6}) is the Legendre transformation of the functional $W(j)$ defined in 
the last line of eq.~(\ref{X2}). 

It remains to show that $W(j)$ agrees with the generating functional of 
connected Green's functions. To see this, recall that the Gibbs measure eq.~(\ref{1.1}) 
is the unique solution of the unconstrained variation principle eq.~(\ref{1.2}), 
irrespective of the actual form of the action. Since we have an additional linear 
term in the action for the $\mu$-minimisation in the definition of $W(j)$ eq.~(\ref{X2}), 
the solution of this $\mu$-minimisation must be a modified Gibbs measure with the 
additional linear term in the action,
\begin{align}
d\mu_j(\phi) &= Z(j)^{-1}\,d\phi\,\exp\Big\{- \hbar^{-1} \left(\, S(\phi) -
(j,\phi)\,\right) \Big\} \label{A4}
\\[2mm]
Z(j) &=  \int d\phi \,\exp\Big\{- \hbar^{-1} \left(\,
S(\phi) -(j,\phi)\,\right) \Big\}\,.
\end{align}
The corresponding value of the minimum in eq.~(\ref{X2}) is then given by the modified 
partition function with a linear term in the action, 
\begin{align}
-W(j)&\equiv& \mathop{\mbox{inf}}_\mu\Big[\,\langle \, S - (j,\phi) \,\rangle_{\mu} 
- \hbar \,\As(\mu)\,\Big] =
\langle \, S - (j,\phi) \,\rangle_{\mu_j} - \hbar \,\As(\mu_j) = 
- \hbar \, \ln Z(j)\,. \label{A12}
\end{align}
Clearly, this identifies $W(j)$ as the usual generating functional of connected 
Green's functions and hence the Legendre transform eq.~(\ref{X4}) as the generator
of 1PI proper functions. The inverse Legendre transformation
\begin{align}
W(j) = \mathop{\sup}_\varphi \,\Big[\, (j,\varphi) - \Gamma(\varphi) \,\Big]
\label{X5}
\end{align}
implies $\delta \Gamma / \delta \varphi(x) = j_\varphi(x)$. Finally, 
eqs.~(\ref{X4}) and (\ref{X5}) entail that both functionals $W(j)$ and $\Gamma(\varphi)$ must 
be \emph{convex}.

\section{The variation principle for $\phi^4$ theory}
\label{app:B}
In this appendix, we will sketch the application of the variation 
principle to $\phi^4$ theory. This is a standard application of the 
variational method and dates back at least to Ref.~\cite{Stevenson:1985zy}, but we 
will repeat it here within our formulation to clearify the distinction between 
the functional and  linear response approach. 
We start with the Lagrangian (in Euclidean space)
\begin{equation}
\mathscr{L} = \frac{1}{2}( \partial_\mu \phi)^2 +
\frac{m^2}{2}\,\phi^2 + \frac{\lambda}{4!} \,\phi^4\,.
\end{equation}
Our trial measures will be Gaussians characterised by a variation kernel $\omega(x,y)$,
\begin{equation}
\rho(\phi) \sim \exp \left[ -\frac{1}{2} \int \d x \d y 
\left(\phi(x) -\varphi(x)\right)\omega (x,y)\left(\phi(y) -\varphi (y)\right) \right]\,.
\end{equation}
We have centered the Gaussian at the classical field $\varphi(x)$ in order to 
obey the constraint $\langle \phi \rangle = \varphi$ for the \emph{linear response} approach;
in the \emph{functional} approach based on the unconstrained free action, we can set 
$\varphi = 0$, cf.~section \ref{sec:2}. Next, we compute the average action given by
\[
\langle \, S \, \rangle_\mu =
\int \d[d] x\, \left[ \,\frac{1}{2}\left(-\Box_x + m^2\right)
\,\langle\,\phi(x)^2 \,\rangle_\mu
 + \frac{\lambda}{4!} \,\langle\,\phi(x)^4\,\rangle_\mu \,\right]\,.
\]
The relevant correlators can easily worked out using Wick's theorem
\begin{align}
\langle\,\phi(x)\,\rangle_\mu &= \varphi(x) \nonumber \\[2mm]
\langle\,\phi(x)\,\phi(y)\,\rangle_\mu &=
\varphi(x)\,\varphi(y) + \hbar\omega^{-1}(x,y) \nonumber \\[2mm]
\langle\,\phi(x)^4\,\rangle_\mu &= \varphi(x)^4 + 6 \hbar \,\omega^{-1}(x,x)\,
\varphi(x)^2 + 3\hbar^2 \,\omega^{-2}(x,x)\,. \label{B2}
\end{align}
The average action thus takes the form\footnote{The action of the Laplace operator
is understood as \quad
$\displaystyle \Box_x\,\omega^{-1}(x,x) \equiv \partial_\mu^x\,\partial_\mu^y\,
\omega^{-1}(x,y)\Big\vert_{y=x}$\,.}
\begin{align}
\langle \,S\,\rangle_\mu &= S(\varphi) + \frac{\hbar}{2}\,\int \d^d x\,
\left[ - \Box_x + m^2 + \frac{\lambda}{2}\,\varphi(x)^2 \right]\,\omega^{-1}(x,x) 
 + \frac{\hbar^2 \lambda}{8}\,\int \d^dx\,\omega^{-2}(x,x) \,.
\label{B4}
\end{align}
The calculation of the entropy is slightly more involved. From the explicit
form of the Gaussian measure and the definition, eq.~(\ref{1.3}),
we have formally
\begin{eqnarray}
\As(\mu) &=& \langle\, - \ln \rho \,\rangle_\mu
= \int d\mu(\phi)\,\Big[\,\ln Z + \frac{1}{2\hbar} \int \d^d(x,y)\,(\phi(x)-\varphi(x))
\,\omega(x,y)\,(\phi(y)-\varphi(y))\,\Big] \nonumber \\[2mm]
&=& \frac{1}{2}\,\ln\det \left(\frac{\omega}{2 \pi\hbar}\right) + \nonumber
\\[2mm]
&& {}+ \frac{1}{2\hbar} \int \d^d(x,y)\,\Big[\, \omega(x,y)\,\langle\,\phi(x)\,
\phi(y)\,\rangle_\mu - 2 \varphi(x)\omega(x,y) \,\langle\,\phi(y)\,\rangle_\mu +
\varphi(x)\, \omega(x,y)\,\varphi(y)
\,\Big]\,. \nonumber
\end{eqnarray}
(An implicit regularisation is understood.) Using Wick's theorem for the correlators,
all dependence on the classical field $\varphi$ drops out,\footnote{This is 
expected since the entropy encodes the available phase space for quantum fluctuations,
which is not affected by the field shift $\varphi$.}
\begin{equation}
\As(\mu) = \frac{1}{2} \,\mathrm{Tr}\,\left\{\, \mathbbm{1}
- \ln \left(\frac{\omega}{2 \pi\hbar}\right)\,\right\}\,.
\label{B5}
\end{equation}
The free action (\ref{1.2}) for the Gaussian measure in $\phi^4$ theory is now
$F(\omega,\varphi) = \langle \, S \,\rangle_\mu - \hbar \,\As(\mu)$ with 
eqs.~(\ref{B4}) and (\ref{B5}) providing the details. 
At this point, the functional and linear response approach start to deviate.

\subsection{The functional approach}
As explained in section \ref{sec:2}, we simply take the \emph{unconstrained}
free action $F(\omega,\varphi=0)$ and vary w.r.t.~the kernel $\omega$, or 
rather, its inverse. After a short calculation, one finds
\begin{align}
\frac{\delta F(\omega,\varphi=0)}{\delta \omega^{-1}(x,y)} =  
\frac{\hbar}{2} \left[- \Box_x + m^2 
+ \frac{\hbar \lambda}{2}\, \omega^{-1}(x,y) \right]\,\delta(x,y) - 
\frac{\hbar}{2}\omega(y,x) \stackrel{!}{=} 0\,.
\label{B6}
\end{align}
Due to the absence of a classical field, the solution of this \emph{gap equation}
in the functional approach has global translation and rotation invariance, 
i.e.~the kernel has the Fourier representation
\begin{align}
\omega(x,y) = \int \frac{\d^d k}{(2\pi)^d}\,e^{-i k \cdot (x-y)}\,\omega(k)\,,
\label{B7a}
\end{align}
where $\omega(k)$ depends only on the modulus $k = |k_\mu|$.
The gap equation in momentum space takes the simple form
\begin{equation}
\omega(k) = k^2 + m^2 + \frac{\hbar \lambda}{2}\,\int \frac{\d^d p}{(2\pi)^p}\,
\frac{1}{\omega(p)}\,.
\label{B8}
\end{equation}
The solution is the covariant dispersion relation for a \emph{free boson} of
a dynamically generated mass,
\begin{equation}
\ww(k) = k^2 + M^2\,,
\label{B9}
\end{equation}
where $M$ is implicitly determined by the non-linear equation
\begin{equation}
M^2 = m^2 + \frac{\hbar\lambda}{2} \int \frac{\d^d k}{(2\pi)^d}\,\frac{1}{k^2 +
M^2}\,.
\label{B10}
\end{equation}
Finally, we can now compute all connected Green functions of the solving Gaussian 
measure, construct a generating functional $W(j)$ and Legendre transform to obtain the 
effective action $\Gamma(\varphi)$. Obviously, all proper vertices  vanish and we get
\begin{equation}
\Gamma(\varphi) = \int \d^d x\,\left[ \frac{1}{2}\, (\partial_\mu \varphi)^2 + 
 \frac{M^2}{2}\, \varphi^2 \right]\,.
\label{B12}
\end{equation}
It should finally be noted that the $p_0$-integration in the covariant 
mass equation (\ref{B10}) can be performed explicitly. For $d=4$, we obtain
\begin{equation}
M^2 = m^2 + \hbar\,\frac{\lambda}{4} \int \frac{d^3 p}{ (2 \pi)^3}\,
\frac{1}{\sqrt{\vek{p}^2 + M^2}}\,,
\label{B13}
\end{equation}
which coincides exactly with the result of the traditional variational 
approach in the Hamiltonian picture based on a Gaussian wave functional.

\subsection{The linear response approach}
In this case, we have to retain the classical field in order to comply
with the constraint $\langle \phi \rangle = \varphi$. From the variation 
of the \emph{constrained} free action  $F(\omega,\varphi)$, we obtain 
the modified gap equation
\begin{align}
\frac{2}{\hbar}\,\frac{\delta F(\omega,\varphi)}{\delta \omega^{-1}(x,y)} = 
\left[- \Box_x + m^2 + \frac{\lambda}{2}\,\varphi(x)^2 + 
\frac{\hbar \lambda}{2}\, \omega^{-1}(x,y) \right]\,\delta(x,y) - 
\omega(y,x) =  0\,.
\label{B14}
\end{align}
As expected in the linear response approach, the optimal kernel $\omega_\varphi$ 
determined from this equation depends implicitly on the classical field $\varphi$,
while the effective action is  simply $\Gamma(\varphi) = F(\omega_\varphi,\varphi)$. 
Since the external fields are arbitrary, the implicit $\varphi$-dependence of the 
optimal kernel $\omega_\varphi$ spoils the translational or rotational invariance. 
As a consequence, we \emph{cannot} go to momentum space as we did in eq.~(\ref{B7a}) 
within the functional approach, which complicates the solution of eq.~(\ref{B14}) 
considerably. 

One possible approach is to attempt a solution of eq.~(\ref{B14}) by expanding in powers 
of the classical field. Upon comparing the two gap equations (\ref{B6}) and (\ref{B14}),
we conclude that 
\begin{align}
\omega_\varphi = \omega + \mathscr{O}(\varphi^2)\,,
\end{align}
where $\omega$ on the rhs is the translationally invariant solution of 
eq.~(\ref{B6}) found earlier.  If we insert the expansion 
back into the free action, we find the effective action
\begin{equation}
\Gamma(\varphi) = \int \d^d x\,\left[ \frac{1}{2}\, (\partial_\mu \varphi)^2 + 
 \frac{M^2}{2}\, \varphi^2  + \mathscr{O}(\varphi^4)\right]\,.
\label{B12x}
\end{equation}
Notice that the quadratic pieces in $\Gamma$ are identical for the functional 
and linear response approach. This is generally true because the Gaussian ansatz
for our trial measure is already quadratic in the fields, and the kernels of the two 
approaches agree at $\varphi=0$. Thus we arrive at the important conclusion:
\begin{quote}
\emph{For Gaussian trial measures, the functional and linear response approach 
always lead to the same quadratic pieces in the effective action, i.e. to the 
same propagators.} 
\end{quote}
Differences arise in the vertices, i.e.~the higher powers of the classical field:
While all vertices vanish in the functional approach, the linear response 
formulation allows for $\mathscr{O}(\varphi^4)$ pieces in eq.~(\ref{B12x}) even though
we only used Gaussian trial measures. 

To examine the structure of these vertex corrections in more detail, we can restrict
our investigation to \emph{constant} classical fields whence the effective action 
reduces to the \emph{effective potential}. This has the benefit that eq.~(\ref{B12}) 
will again be translationally invariant and can thus be solved in momentum space. 
In fact, we can simply repeat the calculation of the functional approach with the 
replacement $m^2 \to m^2 + \frac{\lambda}{2}\,\varphi^2$. To obtain the effective 
potential, we now have to insert the expression for $\omega_\varphi$ back into 
the free action $F(\omega,\varphi)$. The resulting effective potential is rather 
complicated and cannot be expressed in closed form: 
\begin{align}
U_{\rm eff}(\varphi) = \frac{m^2}{2}\,\varphi^2 + \frac{\lambda}{4!} \,\varphi^4
&+ \int \frac{d^d k}{(2\pi)^d}\,\frac{k^2 + m^2 + \frac{\lambda}{2}\,\varphi^2}
{k^2 + M^2(\varphi)} + \frac{\hbar^2\lambda}{8}\,\left[\,\int \frac{d^d k}{(2\pi)^d}\,
\frac{1}{k^2 + M^2(\varphi)}\,\right]^2 \nonumber \\[2mm]
&+ \frac{\hbar}{2} \int \frac{d^d k}{(2\pi)^d}\,
\ln\big[k^2 + M^2(\varphi)\big]\,,
\label{B17}
\end{align}
where $M^2(\varphi)$ is the solution of 
\begin{align}
M^2(\varphi) = m^2 + \frac{\lambda}{2}\,\varphi^2 + \hbar \,\frac{\lambda}{2}
\int \frac{\d^d k}{(2\pi)^d}\,\frac{1}{k^2 + M^2(\varphi)} \,.                                       
\end{align}

All these fomulae are understood to be properly regularised and subject to 
subsequent renormalisation. (We will not discuss the necessary counter terms 
in more detail.) 

The effective potential (\ref{B17}) has an interesting structure: The first two terms 
are the classical potential \emph{including} the bare $\varphi^4$ vertex, 
while the remaining quantum corrections contain all powers of $\hbar$ and $\lambda$.
When expanding in powers of the fields, we obtain
\begin{align}
U_{\rm eff}(\varphi) = \frac{M^2}{2}\,\varphi^2 + \frac{\lambda}{4!}\,
\Big[1 + \mathscr{O}(\hbar)\Big]\,\varphi^4 + \mathscr{O}(\varphi^6)\,,
\end{align}
where $M^2$ is the same field-independent dynamical mass eq.~(\ref{B10}) as in the 
functional approach. We will not further investigate the vertex corrections and 
their renormalization. 

\section{The gap equtaion in the linear response approach}
\label{app:X}
As mentioned in section \ref{sec:F}, the optimal kernel $\wb_\Ac$ in the linear response 
approach is no longer translationally invariant and has no special colour or 
Lorentz symmetry. As a consequence, we cannot reduce the free action and the ensuing
gap equation to a system for a single-component function $\omega(k)$ in momentum space. Instead, 
we have to work with the complete matrix $\wb^{ab}_{\mu\nu}(x,y)$ in position 
space. The corresponding formulas are listed below for completeness, although they are 
only necessary to compute radiative corrections to higher vertices, which we do not consider.

The average action follows by inserting the correlators (\ref{3.13.1}) into the classical action.  
After some lengthy calculation, we obtain 
\begin{align}
\langle S_{\rm gf} \rangle_\mu =& S_{\rm gf}(\Ac) + 
\frac{1}{2} \int d(x,y)\,\wb_{\mu\nu}^{aa}(x,y)\,\Big[ - \Box_x \,\delta_{\mu\nu} + 
(1- \xi^{-1})\,\partial_\mu^x \partial_\nu^x\,\Big]\,\delta(x,y)
\nonumber \\[1mm]
&+g \,f^{abc}\,\int d(x,y) \delta(x,y)\,\partial_\mu^x\,\Big\{ [\wb^{-1}]^{ab}_{\nu\mu}(x,y)\,
\Ac_\nu^c(y) + [\wb^{-1}]^{ac}_{\nu\nu}\,\Ac_\mu^b(y) + [\wb^{-1}]^{bc}_{\mu\nu}\,\Ac_\nu^a(x)\Big\}
\nonumber \\[1mm]
&+ \frac{g^2}{4} f^{cab}\,f^{cde}\,\int dx\,\Bigg\{ 
[\wb^{-1}]^{ab}_{\mu\nu}(x,x)\,[\wb^{-1}]^{de}_{\mu\nu}(x,x) + 
[\wb^{-1}]^{ad}_{\mu\mu}(x,x)\,[\wb^{-1}]^{be}_{\nu\nu}(x,x)  
\nonumber \\[1mm]
&  \hspace*{35mm}+
[\wb^{-1}]^{ae}_{\mu\nu}(x,x)\,[\wb^{-1}]^{bd}_{\nu\mu}(x,x)
\nonumber \\[1mm]
&  \hspace*{35mm}+ 
[\wb^{-1}]^{ab}_{\mu\nu}(x,x)\,\Ac_\mu^d(x)\Ac_\nu^e(x) +
[\wb^{-1}]^{ad}_{\mu\mu}(x,x)\,\Ac_\nu^b(x)\Ac_\nu^e(x)
\nonumber \\[1mm]
&  \hspace*{35mm}+
[\wb^{-1}]^{ae}_{\mu\mu}(x,x)\,\Ac_\nu^b(x)\Ac_\mu^d(x) +
[\wb^{-1}]^{bd}_{\nu\mu}(x,x)\,\Ac_\nu^a(x)\Ac_\mu^e(x)
\nonumber \\[1mm]
&  \hspace*{35mm}+ 
[\wb^{-1}]^{be}_{\nu\mu}(x,x)\,\Ac_\mu^a(x)\Ac_\mu^d(x)+ 
[\wb^{-1}]^{de}_{\mu\nu}(x,x)\,\Ac_\mu^a(x)\Ac_\nu^b(x) \,\Bigg \}\,.
\label{X:1} 
\end{align}
To find the free action, we have to subtract the relative entropy, cf.~eqs.~(\ref{3.17})
and (\ref{3.24}),
\begin{align}
\Asb &= \frac{1}{2}\,\Tr\Big\{ \mathbbm{1} - \ln\left(\frac{\wb}{2\pi}\right)\Big \}  + 
\langle \ln\mathscr{J}\rangle_{\wb, \Ac} 
\nonumber \\[2mm]
&= \frac{1}{2}\,\Tr\Big\{ \mathbbm{1} - \ln\left(\frac{\wb}{2\pi}\right)\Big \} 
- \frac{1}{2}\int d(x,y)\,\chi^{ab}_{\mu\nu}(x,y)\,\Big[
\Ac_\mu^a(x)\Ac_\nu^b(y) + [\wb^{-1}]^{ab}_{\mu\nu}(x,y) \Big]\,.
\label{X:2}
\end{align}
It is now straightforward, though very cumbersome, to derive the gap equation by minimising 
$F = \langle S \rangle_\mu - \Asb$ w.r.t.~the kernel $\wb$, or rather its inverse. The result can 
be put in the form
\begin{align}
\wb^{ab}_{\mu\nu}(x,y) & = \delta^{ab}\,\Big[ - \Box_x \,\delta_{\mu\nu} + (1- \xi)^{-1}\,\partial_\mu^x
\partial_\nu^x\Big]\,\delta(x,y) + \chi^{ab}_{\mu\nu}(x,y) +  
\nonumber \\[1mm]
& + g^2 \,\Big\{ \,f^{cab}\,f^{cde}\,[\wb^{-1}]^{de}_{\mu\nu}(x,x) + 
f^{cad}\,f^{cbe}\,\left( \delta_{\mu\nu}\,[\wb^{-1}]^{de}_{\alpha\alpha}(x,x) - 
[\wb^{-1}]^{de}_{\mu\nu}(x,x)\right) \,\Big\}\,\delta(x,y)
\nonumber \\[1mm]
& + 2g\,f^{abc}\,\Big\{\,(\partial_\mu \Ac_\nu^c(x)) - \partial_\nu \Ac_\mu^c(x) + 
\partial_\alpha \Ac_\alpha^c(x) \,\Big \}\,\delta(x,y)
\nonumber \\[1mm]
&+ g^2\,\Big\{\, f^{cab}\,f^{cde}\,\Ac_\mu^d(x)\Ac_\nu^e(x) + f^{cad}\,f^{cbe}\,\delta_{\mu\nu}
\Ac_\alpha^d(x) \Ac_\alpha^e(x)\,\Big\}\,\delta(x,y)\,.
\label{X:3}
\end{align}
Notice that the derivatives in the third line also act on the $\delta$-function, except for the 
first term in the curly brackets, and the curvature also enters with its full colour and Lorentz 
structure.

\medskip\noindent
Upon setting $\Ac = 0$, only the first two lines on the rhs of eq.~(\ref{X:3}) survive.
This is just the position space equivalent to the gap equation (\ref{3.30}), to which 
it reduces when the symmetric form eq.~(\ref{3.6}) is assumed for $\wb$ and $\chi$. 
An exact solution of eq.~(\ref{X:3}) combined with the full curvature equation is 
not feasable. However, the $\Ac$-dependent terms on the rhs are independent of 
the kernel $\wb$ and thus act like an inhomogenity. This suggests and iterative solution
by expanding in powers of the classical field. When inserted back into the free action, 
the $\mathscr{O}(\Ac^2)$ correction to $\wb$ gives rise to vertices up to $\mathscr{O}(\Ac^4)$ 
etc. Thus, the computation of the vertex corrections should at least be feasable for the 
lowest non-trivial orders. 
\end{appendix}

\bibliographystyle{apsrev4-1}
\bibliography{var}
\end{document}